
\input harvmac
\newcount\figno
\figno=0
\def\fig#1#2#3{
\par\begingroup\parindent=0pt\leftskip=1cm\rightskip=1cm\parindent=0pt
\baselineskip=11pt
\global\advance\figno by 1
\midinsert
\epsfxsize=#3
\centerline{\epsfbox{#2}}
\vskip 12pt
{\bf Fig. \the\figno:} #1\par
\endinsert\endgroup\par
}
\def\figlabel#1{\xdef#1{\the\figno}}
\def\encadremath#1{\vbox{\hrule\hbox{\vrule\kern8pt\vbox{\kern8pt
\hbox{$\displaystyle #1$}\kern8pt}
\kern8pt\vrule}\hrule}}

\overfullrule=0pt

%
\def\tilde{\widetilde}
\def\Z{{\bf Z}}
\def\bar{\overline}

\font\zfont = cmss10 

\def\bigone{\hbox{1\kern -.23em {\rm l}}}
\def\ZZ{\hbox{\zfont Z\kern-.4emZ}}

\Title{hep-th/9507050, HUTP-95/A023, IASSNS-HEP-95-58}
{\vbox{\centerline{Dual String Pairs }
\bigskip
\centerline{With $N=1$ and $N=2$ Supersymmetry}
\bigskip
\centerline{ in Four Dimensions}}}
\smallskip
\centerline{Cumrun Vafa}
\smallskip
\centerline{\it Lyman Laboratory of Physics}
\centerline{\it Harvard University}
\centerline{\it Cambridge, MA 02138}
\smallskip
\centerline{Edward Witten}
\smallskip
\centerline{\it School of Natural Sciences, Institute for Advanced Study}
\centerline{\it Olden Lane, Princeton, NJ 08540, USA}\bigskip

\medskip

\noindent
%
Based on a simple adiabatic argument and by considering the
heterotic string counterpart of certain symmetries of Type II superstrings
such as $(-1)^{F_L}$ and orientation reversal, we construct orbifold candidates
for dual pairs of heterotic and Type II string theories with $N=2$ and $N=1$
supersymmetry.  We also analyze from a similar point of view the ${\rm K3}$
fibrations that enter in recently proposed $N=2$ candidates and use this
structure together with certain orientation-reversing symmetries
to construct $N=1$ dual pairs.  These pairs involve generalizations
of Type I vacua which can be equivalent to $E_8\times E_8$
heterotic strings, while standard Type I vacua are related to $SO(32)$.

\Date{July, 1995}

\newsec{Introduction}

One of the recent lessons  in physics is that there is
apparently only one string theory.   $E_8\times E_8$ and $SO(32)$
heterotic strings and  Type IIA, Type IIB, and Type I
superstrings are apparently all manifestations of one underlying theory,
which also has the well-known wild card, eleven-dimensional supergravity,
as one of its limits.

\nref\sen{J.H. Schwarz and A. Sen,``Duality Symmetries of 4d Heterotic
Strings'', Phys. Lett. {\bf B312} (1993) 105.}
\nref\hull{C.M. Hull and P.K.  Townsend, ``Unity of Superstring
Dualities'', hep-th/9410167.}
\nref\ewitten{E. Witten ``String Theory Dynamics In Various Dimensions,''
hep-th/9503124, to appear in Nucl. Phys. B.}
String theory dualities leading to such statements
have originally been studied in models with
$N=4$ and $N=8$ supersymmetry
 \refs{\sen - \ewitten}.
In field theory, the physical phenomena
controlled by electric-magnetic duality tend to become richer
as the amount of supersymmetry is reduced (for a recent review see
\ref\seiberg{N. Seiberg, ``The Power Of Duality: Exact
Results in 4d SUSY Field Theory,'' hep-th/9506077.}).
\nref\kachruvafa{S. Kachru and C. Vafa,``Exact Results for $N=2$
Compactifications of Heterotic Strings'', hep-th/9505105, to appear
in Nucl. Phys. B.}
\nref\fhsv{S. Ferrara, J. A. Harvey, A. Strominger, and C. Vafa,
``Second-Quantized Mirror Symmetry'', hep-th/9505162.}
Recent developments involving string-string duality applied to models
with $N=2$ supersymmetry in four dimensions \refs{\kachruvafa,\fhsv}
have shown that this is true in string theory as well.
(For tests of some of the models in \kachruvafa\
see \nref\lerche{A. Klemm,
W. Lerche and P. Mayr,``K3-Fibrations and Heterotic-Type II String
Duality'', hep-th/9506112.}\nref\klt{V. Kaplunovsky, J. Louis,
and S. Theisen,``Aspects of Duality in N=2 String Vacua,''
hep-th/9506110.}\nref\nar{I. Antoniadis,
E. Gava, K. Narain and T. Taylor,
unpublished.} \refs{\lerche - \nar}.)

The duality between string theories has an amazing
problem-solving ability: a problem of dynamics defined by writing
down one string theory can sometimes be solved by finding a dual theory
to which it is equivalent.  In that respect, it is fascinating to
note that the heterotic string is naturally chiral, with non-abelian
gauge groups, like the real world at high energies, while Type II strings
are naturally non-chiral, with abelian   gauge groups, like the real
world at low energies.  It is roughly as if, in the real world, the heterotic
string should be used to describe grand unification and an equivalent
Type II description should be used to understand the physical properties
of the low energy vacuum.

The main purpose of the present paper is to construct additional
dual pairs in four dimensions, with $N=2$ or $N=1$ supersymmetry; by a dual
pair we mean simply a pair of equivalent Type II and heterotic string vacua.
Such dual pairs have particular interest in the $N=1$ case, where one
might meet models similar to the real world (though the models we will
actually explore in this paper will not be terribly realistic) and one
might encounter issues of supersymmetry breaking and the cosmological constant.
Our models will be constructed by taking certain orbifolds of models
-- such as those of \kachruvafa\ and \fhsv\ -- for which dualities have
been found.
 This must be done with some care, because in general
orbifolding seems not to  commute with string-string duality; that is,
it seems that in general  the
orbifold of a dual pair is not a dual pair.

This paper is organized as follows.  In section two, we discuss
the inequivalent worldsheet structures by which $N=2$ and $N=1$
space-time supersymmetry can be realized by Type II superstrings in four
dimensions.  We illustrate the discussion with a few simple examples of what
appear to be dual pairs with $N=1$ or $N=2$ supersymmetry.  We also make,
in a similar vein, some simple remarks relevant to new $N=4$ models
\nref\chl{S. Chaudhuri, G. Hockney, and J. Lykken,``Maximally
Supersymmetric String Theories in D$<$10,'' hep-th/9505054; S.
Chaudhuri, Talk given at String'95, USC, March 95, to appear in
the  proceedings.}
\nref\polchinski{S. Chaudhuri and J. Polchinski,``Moduli Space
of CHL Strings,'' hep-th/9506048.}
\refs{\chl - \polchinski} which can also be used as starting points
for constructing dual pairs with $N<4$ supersymmetry.
In section three, we try to deduce the dualities discovered in
\kachruvafa\ from six-dimensional string-string duality
by using the interpretation of the relevant Calabi-Yau
manifolds as fiber bundles with K3 fibers \lerche, by means of which
the $j$-functions appearing
in the examples of \kachruvafa\ have been  related to the
results of Lian and Yau \ref\lya{B.H. Lian and S.-T. Yau,
``Arithmetic Properties of the Mirror Map and Quantum
Coupling'', hep-th/9411234.} on families of ${\rm K3}$'s.
Though we do not succeed
in fully explaining the models of \kachruvafa , we draw a few lessons that we
then use
to construct additional dual pairs with $N=1$ supersymmetry
-- at least one such model for each $N=2$ model in \kachruvafa.
These models involve a generalization of Type I superstrings, and
have the notable property of reproducing certain $E_8\times E_8$
heterotic string vacua, while standard Type I models are limited
to $SO(32)$.

\newsec{World-Sheet Structure And Space-Time Supersymmetry}

For a heterotic string vacuum in four dimensions to have $N=1$
space-time supersymmetry, the world-sheet theory must be a
$(0,2)$ superconformal field theory with a $U(1)$ $R$ symmetry with
integral charges \ref\inte{A. Sen, ``(2, 0) Supersymmetry
and Space-time Supersymmetry in the Heterotic String Theory,''
Nucl. Phys. {\bf B278} (1986) 289;``Heterotic String Theory on
Calabi-Yau Manifolds in the Green-Schwarz Formalism,''
Nucl. Phys. {\bf B284} (1987) 423\semi T. Banks, L.J. Dixon, D. Friedan
and E. Martinec,``Phenomenology and Conformal Field Theory or Can
String Theory Predict the Weak Mixing Angle,''
Nucl. Phys. {\bf B299} (1988) 613.}.
  Likewise, $N=2$ space-time supersymmetry
for the heterotic string comes from a $(0,4)$ superconformal  world-sheet
theory with an $SU(2)$ $R$ symmetry \ref\badi{T. Banks and
L.J. Dixon,``Constraints on String Vacua with Space-Time Supersymmetry,''
Nucl. Phys. {\bf B307} (1988) 93.}.
For Type II superstrings, there are instead several distinct world-sheet
structures that lead to a given space-time supersymmetry.  For $N=2$
there are the following possibilities.

(1)  $N=2$ space-time supersymmetry in four dimensions
can arise from a $(2,2)$ world-sheet
theory with left and right moving $U(1)$ $R$ symmetries.  This is the
most familiar case; it  arises in compactification of Type II superstrings
on a Calabi-Yau manifold.   One space-time supersymmetry is carried
by left-moving degrees of freedom on the world-sheet and one is carried
by right-moving degrees of freedom.

(2) $N=2$ space-time supersymmetry can likewise arise from a $(1,4)$
world-sheet theory.  In this case, both space-time supersymmetries
are carried by right-moving degrees of freedom on the world-sheet.

(3) Finally, one can consider a left-right symmetric world-sheet structure
with $(4,4)$ world-sheet supersymmetry in which by introducing in some
way unorientable world-sheets one projects the massless states onto
states that are left-right symmetric.  (This can be done by introducing
open strings as in the Type I theory, or  by constructions -- discussed
later -- that are similar to the orientifolds of \ref\dlp{J. Dai, R. Leigh,
and J. Polchinski, ``New Connections Between String Theories,'' Mod. Phys.
Lett.
{\bf A4} (1989) 2073; R. Leigh, ``Dirac-Born-Infeld Action From Dirichlet
Sigma Model,'' Mod. Phys. Lett. {\bf A4} (1989) 2767.}.)
In this case, both space-time supersymmetries
are carried by linear combinations of left and right-moving variables.

Similarly, two distinct world-sheet structures give $N=1$ supersymmetry
in four dimensions for Type II superstrings.  They are analogs
of cases (2) and (3) above.

($2'$) $N=1$ supersymmetry can arise from a $(1,2)$ superconformal world-sheet
theory with a right-moving $U(1)_R$ symmetry.  The supersymmetry is carried
by right-movers.

($3'$) $N=1$ supersymmetry can likewise arise from a $(2,2)$ world-sheet
with a projection on left-right invariant states.  The structure arising
generalizes that of a Type I theory compactified on
a Calabi-Yau manifold:
the $N=1$ space-time supersymmetry is carried by a
combination of left-moving and right-moving degrees of freedom.

In the rest of this section, we first answer a question of basic
importance for $N=2$ strings in four dimensions: where is the dilaton?
For method (1) of realizing space-time supersymmetry, the dilaton is
known to be
part of a hypermultiplet \ref\seiob{N. Seiberg,``Observations
On the Moduli Space of Superconformal Field Theories,'' Nucl. Phys.
{\bf B303} (1988) 286.}; in case (2), we will see
that the dilaton is part of a vector multiplet while in case (3) it is
a linear combination of part of a vector multiplet and part of a
hypermultiplet.  Because (except for gauge couplings) vector multiplets
and hypermultiplets are decoupled at low energies in $N=2$ theories
in four dimensions \ref\vanp{B. de Wit, P. Lauwers and A. Van Proeyen,
``Lagrangians Of $N=2$ Supergravity-Matter Systems,''
Nucl. Phys. {\bf B255} (1985) 269.}, it follows from the above that
 in case (1), there are no quantum
corrections to the vector moduli space - a fact that has played an important
role recently -  while in case (2), there are
no corrections to the hypermultiplet moduli space, and in case (3)
(as the dilatons can couple at low energy to modes of both kinds)
there can be corrections to each moduli space.

After locating the dilaton, we  move on to discuss
simple illustrative models of types (2) and (3), and also of types
($2'$) and ($3'$), for which heterotic string duals can apparently be
identified.  Finally,
we discuss in a somewhat similar spirit new $N=4$ models
\chl\
that can apparently be modified to give additional dual pairs with $N<4$.

\subsec{Finding The Dilaton}

First we recall how the dilaton arises in string theory in general.
Left-moving massless states have helicity at most one in absolute
value; likewise the right-moving states have helicity at most one.
By taking the tensor product of left and right-moving massless
states, one can make the graviton, of helicity $\pm 2$.  One can also,
by taking the tensor product of a left-moving state of helicity $-1$ with
a right-moving state of helicity $1$, construct a scalar; a second scalar
comes from the tensor product of left-moving helicity $1$ and right-moving
helicity $-1$.  The symmetric combination of these states is known
as the dilaton.  (For oriented closed strings, the antisymmetric combination
of these states is the axion; for unoriented superstrings, the axion
is instead a Ramond-Ramond state.)

Irreducible massless $N=2$ representations have four  states of helicities
 $(j,j-1/2,j-1/2,j-1)$ where $j$ is an integer or half-integer which
is the ``highest weight'' of the representation.
Because of constraints such as CPT, pairs of irreducible representations
combine into the $N=2$ multiplets that actually appear in Lagrangians.
For instance, the $N=2$ vector
multiplet is the direct sum   of representations with highest weight $j=1$
and $j=0$ (which contain respectively the vector states of helicity
$1$ and $-1$), while the hypermultiplet is the direct sum of two
$N=2$ multiplets with highest weight $j=1/2$.

Now it is fairly straightforward, by locating the left and right-moving
states of helicity $\pm 1$ in appropriate supermultiplets and taking the
tensor products, to see what
kind of supermultiplet contains the dilaton.  We consider the three cases
in turn.

(1) For  (2,2) models, the left-moving states and right-moving states
each carry $N=1$ supersymmetry -- leading to $N=2$ for the tensor product.
Irreducible massless multiplets of $N=1$ have precisely two helicity
states.  The left-moving state of helicity $-1$ is thus in a multiplet
containing two states of helicities $(-1/2,-1)$, and the right-moving
state of helicity $1$ is in a multiplet containing two states of helicities
$(1,1/2)$.  Taking the tensor product of these $N=1$ multiplets
gives an $N=2$ multiplet $W$ with four states of helicities $(1/2,0,0,-1/2)$.
which we recognize as  part of a hypermultiplet.
Therefore, the dilaton is part of a hypermultiplet when $N=2$ supersymmetry
is realized using a (2,2) world-sheet structure.

(2) If $N=2$ supersymmetry is realized in a (1,4) model, then the
left-moving states carry Lorentz quantum numbers only and the right-moving
states form $N=2$ multiplets.  To find the dilaton, we consider the tensor
product of a left-moving state of helicity $-1$ with a right-moving
$N=2$ multiplet of helicities $(1,1/2,1/2,0)$.  The tensor product
gives an irreducible $N=2$ representation with states of helicity
$(0,-1/2,-1/2,-1)$, which we recognize as part of a vector multiplet.
  Hence in this case the
dilaton is in a vector multiplet.  (Precisely the same argument shows that
for $N=2$ heterotic strings, the dilaton is in a vector multiplet; all that
matters is that in each case, the supersymmetry is carried entirely by
right-movers.)

(3) More complex is the case of a world-sheet $(4,4)$ model
with a projection on symmetric combinations of left and right-moving
states.  In this case, the left and right-moving states are each
in $N=2$ multiplets.  We start with  left-moving $N=2$ representations
$A_L$ of highest weight $j=0$, that is
helicities  $(0,-1/2,-1/2,-1)$
and $B_L$ of highest weight
 $j=1$, that is helicities
$(1,1/2,1/2,0)$, and right-moving   $N=2$ multiplets $A_R$ and $B_R$
also        of highest weights 0 and 1, respectively.
The dilaton and axion are contained in the tensor product $A_L\otimes B_R$.
($A_R\otimes B_L$ need not be considered separately, since the projection
on left-right symmetric combinations effectively identifies $A_R\otimes B_L$
with $A_L\otimes B_R$.)  The $N=2$ representation $A_L\otimes B_R$ has sixteen
states and can be decomposed as a direct sum of four irreducible
four-dimensional representations of $N=2$.  Let us find the highest
weights of these irreducible representations.  One of them is the highest
weight state $|0,1\rangle$ of $A_L\otimes B_R$.  Having helicity
one, it is the highest weight of a representation
that is part of a vector multiplet.  To find the
other highest weights, we let $Q'_i,\,\,i=1,2$ be the left-moving
helicity lowering operators, $Q''_i,\,\,i=1,2$ the     right-moving
helicity lowering operators,  $Q_i=Q'_i+Q''_i$ the diagonal combination
that is        observed as the physical $N=2$ supersymmetry, and
$\tilde Q_i=Q'_i-Q''_i$.  Of the states of helicity $1/2$, clearly
$Q_i|0,1\rangle$ are in the vector multiplet headed by $|0,1\rangle$,
and $\tilde Q_i|0,1\rangle$ are highest weights of new $N=2$ representations.
These states have $j=1/2$ and so are part of a hypermultiplet.
At helicity zero we have the following states:
\eqn\jurry{\eqalign{ \epsilon^{ij}Q_iQ_j&|0,1\rangle \cr
                      Q_i\tilde Q_j &|0,1\rangle \cr
                      \epsilon^{ij}\tilde Q_i\tilde Q_j&|0,1\rangle .\cr}}
The state in the first row, being obtained by lowering of $|0,1\rangle$,
is in a vector multiplet; the states in the second row, being obtained
by lowering of the $j=1/2$ highest weight $\tilde Q_j|0,1\rangle$, are
in a hypermultiplet; and the state in the last row, being a $j=0$ highest
weight (of an $N=2$ representation that also contains a state of helicity
$-1$), is in a vector multiplet again.  Now to locate the dilaton,
we note simply that the dilaton of this model is the state obtained
by tensoring a left-moving state of helicity $-1$ with a right-moving
state of helicity $1$ (because of the left-right projection, this is
equivalent to the tensor product of helicities $1$ and $-1$; the axion
is a certain Ramond-Ramond state).  The dilaton is therefore
$D=\epsilon^{ij}Q'_iQ'_j|0,1\rangle$.  Expressing  $D$ as a linear
combination of the states given   in \jurry, we see that for models of
this type, the dilaton is a linear combination of states that belong
to a vector multiplet and states that belong to a hypermultiplet.
\foot{In the orbifold examples discussed later in this section, the $T$ modulus
of the heterotic string is in a vector or hypermultiplet or in a mixture
depending on whether the ${\bf Z}_2$ acts with two, zero, or one
invariant directions on the common ${\bf T}^2$.  This easily verified
statement is converted by string-string duality into the above assertions about
the dilaton.}

We need not repeat this analysis for theories with $N=1$ supersymmetry
in four dimensions, since for $N=1$ there is only one kind of multiplet
-- the chiral multiplet -- containing states of helicity zero.  The
dilaton is therefore in a chiral multiplet in both cases ($2'$) and ($3'$) of
$N=1$ supersymmetry.  There is, however, an interesting general difference
in the way that duality is realized for the two ways of obtaining $N=1$
supersymmetry in four dimensions from Type II.  The Type IIB theory in
ten dimensions has an $SL(2,{\bf Z})$ symmetry mixing the dilaton with
a certain Ramond-Ramond scalar.  Upon compactification to four dimensions,
this turns into a standard $S$-duality for $N=1$ orientifolds (since the
Ramond-Ramond scalar in question becomes the axion). It acts
differently for $N=1$ models with $(1,2)$ world-sheet supersymmetry.

\subsec{A (1,4) Example}

\def\R{{\bf R}}
\def\K3{{\rm K3}}
Our next goal is to construct a simple example of a Type II vacuum with
(1,4) world-sheet supersymmetry and $N=2$ spacetime supersymmetry.
Then we will propose a candidate for the heterotic string dual of this
vacuum.  We start with the Type IIA theory on $\R^6\times \K3$.
This is a familiar
model with $(4,4)$ world-sheet supersymmetry and $N=4$ spacetime supersymmetry
-- two supersymmetries being carried by left-movers and two by right-movers.
The supersymmetries carried by left-movers have one chirality in spacetime
and the supersymmetries carried by right-movers have the opposite space-time
chirality.

We want to modify this model
to obtain a model with only $N=2$ spacetime supersymmetry,
carried entirely by right-movers.  To do  this, we will make an orbifold
projection, dividing by a    symmetry
that commutes with the spacetime supersymmetries that come from right-movers
and anticommutes with the ones that come from left-movers.

The obvious symmetry with that property is $(-1)^{F_L}$, the operation
that acts as $-1$ on states the left-moving part of which is fermionic
and as $+1$ on states the left-moving part of which is bosonic.
Before proceeding, let us ask what $(-1)^{F_L}$ transforms into
under the conjectured duality between the          Type IIA theory on
$\R^6\times \K3$ and the heterotic string on $\R^6\times {\bf T}^4$.
Apart from anything else, the ability to find a symmetry
of the heterotic string with the properties of $(-1)^{F_L}$ is a test of
string-string duality (assuming that $(-1)^{F_L}$ is actually a valid
symmetry of the Type IIA theory with no nonperturbative anomalies).

\def\T{{\bf T}}
\def\S{{\bf S}}
Translating $(-1)^{F_L}$ into heterotic string language can be done easily
as follows.   $(-1)^{F_L}$ acts as $-1$ on all Ramond-Ramond states.
All           24 $U(1)$ vector bosons of the model are Ramond-Ramond
states, so the symmetry must act as $-1$ on all 24 of them.
This means that on the heterotic string side, the symmetry must
be the operator $W=-1$ acting on
the Narain lattice $\Gamma^{20,4}$ describing the
compactification.  Apart from the fact that $W$ transforms the gauge
bosons correctly, the identification of $(-1)^{F_L}$ with $W$
can be checked as follows.  The inversion
of the Narain lattice commutes with the $T$-duality group $SO(20,4;{\bf Z})$,
as expected for $(-1)^{F_L}$.  Also note that, as expected of
$(-1)^{F_L}$, $W$ acts as $+1$ on the two
supersymmetries of one chirality  and as $-1$ on the other two.
 Indeed, for the heterotic string on $\R^6\times {\bf T}^4$,
the inversion $W$ of the Narain lattice acts in particular as $-1$ on $\T^4$;
as in standard constructions of K3 orbifolds, this operation acts
as $-1$ on the spinors of one chirality and as $+1$ on the spinors of the
other chirality.

Now that we have identified the symmetry, we would like to use it
to construct orbifolds with $N=2$ supersymmetry.  To begin with, one
might try to orbifold the Type IIA theory by $(-1)^{F_L}$.  As a way
of obtaining an $N=2$ model, this does not succeed; dividing by
$(-1)^{F_L}$ eliminates two supersymmetries from the untwisted sector,
but one gets two more supersymmetries from the twisted sector, giving in
fact the Type IIB theory on $\R^6\times \K3$.  Incidentally, this
model gives a simple illustration of the fact that orbifolding does not
in general commute with string-string duality.  In
dividing the heterotic string by $W$, one gets a theory whose
twisted sector has 20 antiperiodic
left-moving oscillators.  This gives a $1/4$ mismatch between left- and
right-moving ground state energies, corresponding to
an inconsistent theory with sigma model anomalies. (There is no way
to restore level matching by shifts because -- as all the oscillators
are twisted -- any shift could be absorbed in adding a constant
to the world-sheet bosons.)  So the orbifold of
the Type IIA theory by $(-1)^{F_L}$ is not equivalent to the orbifold
of the heterotic string by $W$.

To modify the construction so that we do get $N=2$ supersymmetry
on the Type II side, we only need to combine $(-1)^{F_L}$ with another
operation in such a way as to avoid getting extra supersymmetry
from the twisted sector.  An obvious way to do this is to consider
the Type IIA theory on $\R^5\times \S^1\times \K3$, and to divide
by a symmetry $Y$ that acts as a $\pi$ rotation of $\S^1$ together with
$(-1)^{F_L}$.  For generic radius of $\S^1$, there are no massless
states in the twisted sector, so we do get a model with      $N=2$
spacetime supersymmetry entirely   carried by right-movers on the world-sheet.

\bigskip
\noindent{\it The Adiabatic Argument And Duality}

We would like to find a heterotic string orbifold that will be equivalent
to the Type II model just described.  First we should ask: Since orbifolding
does not commute with string-string duality in general, why should such a
model exist?  In the particular case at hand, one can give a physical
argument that we consider convincing; we will call it the adiabatic argument.
In this paper we will mostly consider examples to which this
argument applies.

The argument uses the fact that the radius $R$ of $\S^1$ is arbitrary
and that the symmetry $Y$ acts freely on $\S^1$.  If $R$ is very large,
\foot{We want to go to  large $R$ keeping fixed the string
coupling constant in the six-dimensional sense -- a prescription that
is invariant under string-string duality.}
the low energy observer in the Type II theory on $\R^5\times \S^1\times
\K3$ sees a world which -- unless he circumnavigates the $\S^1$ -- is
very hard to distinguish from $\R^6\times \K3$.  Locally, one can use
string-string duality and convert to a heterotic string description on $\R^6
\times \K3$.  Once the equivalence is established locally, the low energy
observer can reasonably expect that it will remain valid globally if
he or she suitably redefines all physical variables in circumnavigating the
$\S^1$.

{}From this argument alone the low energy observer cannot determine precisely
which heterotic string orbifold should correspond to the given Type II
orbifold.  That is because certain distinctions disappear when $R$ becomes
very large.  To see this, consider the Narain lattice $\Gamma^{1,1}$
appropriate to $\S^1$.  We can describe it as consisting of pairs
of integers $(m,n)$ with the inner product of $(m,n)$ and $(m',n')$ being
$mn'+m'n$; this is obviously even and unimodular.  A purely right-moving
momentum vector is of the form $(m,m)$ (for which the inner product
is positive); pure left-moving momenta are $(m,-m)$.  A state with ordinary
spatial momentum and no winding is given by the null vector $(m,0)$ while
winding without momentum would be the null vector $(0,m)$.
Now, let us consider the possible ${\bf Z}_2$ orbifolds of ${\bf S}^1$
(temporarily ignoring questions of level matching which depend on the
coupling to the other degrees of freedom).  A ${\bf Z}_2$ orbifold
is obtained by shifting $\R^{1,1}$ (in which $\Gamma^{1,1}$ is embedded)
by a       vector
$A$ which is one-half of a point in $\Gamma^{1,1}$; moreover, addition
to $A$ of a point in $\Gamma^{1,1}$ will give an equivalent model.
Thus, the non-zero choices of $A$ are essentially $A_1=(1/2,0)$,
$A_2=(1/2,1/2)$, and $A_3=(0,1/2)$.

Let us think about how the resulting theories look to a low energy
observer at large $R$.
The low energy observer can measure ordinary momenta but not windings,
\foot{The winding states of the heterotic string correspond under
string-string duality to soliton states of the Type II theory whose
behavior under $Y$ is not entirely clear.}
and so can measure inner products of $A$ only with ordinary momentum
vectors $(m,0)$.  This means that the shift by $A_1$ looks trivial
to the low energy observer, while the $A_2$ and $A_3$ shifts
look equivalent for large $R$; each of them acts on a momentum $m$
by $(-1)^m$ and so is  equivalent on momentum states to a $\pi$  rotation
of the circle.  Notice that $A_3$ is a null  vector and the shift
by $A_3$ actually is an ordinary rotation, while $A_2$ obeys
$A_2^2=1/2$, and the shift by $A_2$ is not a standard rotation.

The low energy observer thus expects that the orbifold of the Type II
theory by $Y$         will be equivalent to the inversion $W$ of the Narain
lattice together with a shift by $A_2$ or $A_3$.  Which is correct?
This  question can be settled by thinking about level matching.
The transformation $W$ acts as $-1$ on 20 left-moving bosons and so
raises the ground state energy of left-movers by $20/16=5/4$.
The total shift of the left-moving ground state energy will then
be $\Delta_L=5/4+A^2/2$.   As (by supersymmetry) there is no such
shift for right-movers, level-matching requires that $\Delta_L $ should
be an integer  multiple of $1/2$, so we need $A^2=1/2 \,\,{\rm mod}\,
{\bf Z}$.  This fixes $A=A_2$.

A similar effect --
a not purely geometrical shift in the internal $\Gamma^{20,4}$ Narain
lattice
 to preserve level matching for the heterotic string -- was found in a
different example
in \fhsv.  Since in our case the shift is on a geometrical circle
that is part of the six dimensions common to type IIA and heterotic strings,
we have added a somewhat more detailed explanation of why
this does not contradict the experience of the low energy observer.

Let us compactify this model on another circle $\S^1$ down to
four dimensions.  This still gives  a theory with $N=2$.
For later reference, let us summarize the construction of the heterotic
string version of this model.  We have first considered toroidal
compactification down to 6 dimensions, which is characterized
by a Narain lattice $\Gamma^{20,4}$.  We have then compactified
on another two dimensional torus characterized by a Narain lattice
$\Gamma^{2,2}$.  Write a general momentum and winding state as
$$|\gamma^{20,4},\gamma^{2,2}\rangle  \in
\Gamma^{20,4}\oplus \Gamma^{2,2}.$$
Then the ${\bf Z}_2$ action by which  we have modded out is generated by
\eqn\ofact{g|\gamma^{20,4},\gamma^{2,2}\rangle = {\rm exp}(2\pi i
A_2\cdot \gamma^{2,2}) |-\gamma^{20,4},\gamma^{2,2}\rangle}
(where $A_2$ is the vector introduced before but now embedded in
 $\Gamma^{2,2}$), whereas on the Type IIA
side we have modded out by $(-1)^{F_L}{\rm exp} (2 \pi i A_3\cdot
\gamma^{2,2})$.

Let us attempt to check the duality between the
Type IIA and heterotic string   descriptions
in this example.  To see that both theories have the same
massless spectrum,   all we have to check is the untwisted
sector because all the twisted states are massive:  On the Type II side
we have a left-right symmetric shift which thus gives positive mass; on
the heterotic side, even though the shift is asymmetric and by itself
will not preclude a massless state at special radius,
the fact that we have 20 half-integral
oscillators means that the left-moving energy is at least $1/4 >0$.

It is easy to enumerate the massless
states we get from the untwisted sector
after projecting by the ${\bf Z}_2$ action \ofact .  First we consider
vector multiplets.
Both models have
 generically a gauge symmetry $U(1)^4$ coming
entirely from the $\T^2$, so there are three vector multiplets,
with the fourth $U(1)$ vector field
being part of the $N=2$ gravity multiplet.    The
scalar components of the  three
 vector multiplets are the fields usually denoted by $S,T,U$, where
in going from the heterotic to Type IIA we have to exchange
$S\leftrightarrow T$.  The
weak coupling dualities of $S$, $T$ and $U$ will be analyzed at the
end of this section.
Since the dilaton $S$ is also
a member of a vector multiplet -- on both sides, as we have seen --
 the metric on the vector moduli space is expected to be corrected.
Such corrections are certainly present.  For example
if we choose the second circle (the one which brought us down from five
to four
dimensions) to be at the self-dual radius, we get in the heterotic string
description an $SU(2)$ gauge symmetry
at the classical level,  but  quantum mechanically the picture
is different \ref\sw{N. Seiberg and
E. Witten,``Electric-Magnetic Duality, Monopole Condensation, and
Confinement in $N=2$ Supersymmetric Yang-Mills Theory,'' Nucl.
Phys. {\bf B426} (1994) 19.}.
Note that if for $(1,4)$ Type II strings, the
dilaton
were part of a hypermultiplet (as it is in the $(2,2)$ case), we would be
unable to reproduce these corrections on the Type II side and would meet
a contradiction.

There are also
20 hypermultiplets.  Since the dilaton is in a vector multiplet
in each theory, the hypermultiplet moduli spaces are uncorrected,
and for the theories to be equivalent their classical hypermultiplet
moduli spaces must coincide.  This is so; on the Type IIA side,
the moduli space is naturally the moduli space of quantum $\K3$'s, and
for the heterotic string we get the moduli space of Narain lattices
$\Gamma^{20,4}$.
That these coincide is an essential test of string-string
duality in six dimensions.

\subsec{An Orientifold Example}\def\S{{\bf S}}
Our next goal is to construct a simple $N=2$ vacuum of the Type II theory based
on unorientable world-sheets
together with a candidate for its heterotic string dual.

First of all, recall that in the Type IIA superstring theory, the left-movers
on the string world-sheet give spinors of negative chirality in space-time,
and the right-movers give spinors of positive chirality.  The theory is
therefore invariant under reversal of orientation of space-time together with
reversal of the orientation of the world-sheet.  If the target space is
invariant under some orientation-reversing involution, then we get a symmetry
of the string theory with that target space; otherwise we get an equivalence
between a target space and its orientation-reversed counterpart.

Let us make a general comment about such a symmetry $X$.  In general, giving
a classical orientation-reversing  involution of space-time does not quite
determine the operation in string theory.  Though $X^2$ is trivial
as on operation on space-time, it may act as $-1$ on spinors; that is,
one may have $X^2=(-1)^F$.  This situation, however, can always be avoided,
as follows.  One simply sets $\tilde X=(-1)^{F_L} X$ and observes that
(with $X$ being an operation that exchanges left and right-movers) if
$X^2=(-1)^F$
then $\tilde X^2=1$.  So replacing $X$ by $\tilde X$ if necessary, we can
always achieve $X^2=1$.\foot{This discussion is still not precise
enough to fix the action of $X$ on certain black holes, as we will explain
later.}

With this in mind, consider the Type IIA theory compactified on ${\bf
R}^5\times
{\bf S}^1\times \K3$.  If we take the $B$-fields on $\K3$ to vanish,
then the sigma model with this target space is left-right symmetric on the
world-sheet.  This means that the theory is invariant  under a symmetry
$X$ that acts by an orientation-reversing diffeomorphism $x\to -x$ of $\S^1$,
together with exchange of left and right-movers.

What does $X$ correspond to for heterotic strings on $\R^5\times \S^1\times
\T^4$?
Apart from any application we may make of this, the ability to identify
$X$ on the heterotic string side is a test of string-string duality
(unless $X$ is broken explicitly by nonperturbative corrections to the Type IIA
theory).  Obviously, $X$ must correspond to an operation that acts as $x\to -x$
on
$\S^1$.  But what does it do to ${\bf T}^4$?

This question can be answered by using the fact that $X$ must commute
with the appropriate $T$-duality group.  At first sight one may despair
since the center of the usual $T$-duality group $SO(20,4;{\bf Z})$ of $\K3$
has $-1$ as its only non-trivial element, and we have already used this element
in identifying $(-1)^{F_L}$!

However, since we have had to set the $B$-fields to zero to obtain the symmetry
$X$,
the usual conformal field theory moduli space  $SO(20,4;{\bf Z})\backslash
 SO(20,4;{\bf R})/SO(20)\times SO(4)$
 \ref\aspmor{P. Aspinwall
and D. Morrison,``String Theory on K3 Surfaces,'' hep-th/9404151.}\
is replaced by a subspace, which is in fact the classical moduli
space of Einstein metrics on $\K3$
\ref\tod{A. Todorov,``Applications of the Kahler-Einstein-Calabi-Yau
Metric to Moduli of K3 Surfaces,''
 Inv. Math. {\bf 61} (1980) 25.}
$$\left(SO(1,1;{\bf Z}) \backslash
SO(1,1;{\bf R})\right)\times \left(SO(19,3;{\bf Z})\backslash
 SO(19,3;{\bf R})/ SO(19)\times SO(3)\right).$$
The first factor parametrizes the volume of $\K3$ and the second factor
parametrizes Einstein metrics of unit volume.  The decomposition means that
the cohomology lattice has an orthogonal decomposition as $\Gamma^{1,1}\oplus
\Gamma^{19,3}$ and that  with $B$-field zero,
the $T$-duality group can be reduced to the subgroup
$SO(1,1;{\bf Z})\times SO(19,3;{\bf Z})$ that respects this decomposition.

The center of this smaller group now contains the two elements that act
as $+1$ on $\Gamma^{1,1}$ and $-1$ on $\Gamma^{19,3}$, or vice-versa.
Either of these operations gives a symmetry of the theory (they differ
by multiplication by $(-1)^{F_L}$), but if we want $X^2=1$ (rather than
$X^2=(-1)^F$), we must choose $X$ to act as $+1$ on $\Gamma^{1,1}$ and
$-1$ on $\Gamma^{19,3}$.  This can be seen by studying the exchange of
left and right-movers in the Ramond-Ramond sector of the Type IIA theory,
or more simply in the heterotic string description to which we now turn
(where $X^2=(-1)^F$ would arise if $X$ acted with two rather than four
eigenvalues
$-1$ on right-moving bosons).

For an interpretation of $X$ for heterotic strings, it is now clear what we
must
do.
We write the Narain lattice
of ${\bf S}^1\times \T^4$ as $\Gamma^{1,1}\oplus \Gamma^{20,4}=\Gamma^{1,1}
\oplus \Gamma^{1,1} \oplus \Gamma^{19,3}$.  $X$ acts as $-1$ on the first
$\Gamma^{1,1}$ (coming from ${\bf S}^1$), as $+1$ on the second $\Gamma^{1,1}$
(coming from $\T^4$, and as $-1$ on the last factor.  Thus we have succeeded
in finding a heterotic string counterpart of $X$.  Actually, the determination
of $X$ was not quite unique: one could add a translation by half a lattice
vector in the fixed $\Gamma^{1,1}$.  (Such a transformation is a symmetry
of the Type IIA theory that acts only on Ramond-Ramond black holes, so we have
not defined the operation $X$ precisely enough to be able to say if it should
be present here or not.)

It is curious to compare this with the operation $W$ that represents
$(-1)^{F_L}$
for heterotic strings.  $W$ acts on the Narain lattice $\Gamma^{1,1}\oplus
\Gamma^{1,1}\oplus \Gamma^{19,3}$ of $\S^1\times \T^4$ as $+1$ on the first
factor and $-1$ on the last two factors.  This differs from $X$ by exchange
of the first two factors.  Thus, we get the remarkable result that the
 symmetries $(-1)^{F_L}$ and $X$ on ${\bf R}^5\times {\bf S}^1\times \K3$
are in fact conjugate.  One can also easily check, with this description
of $X$ and $W$, that $XW=WX(-1)^F$, as one can anticipate from the Type II
side.

The surprising conjugacy of $X$ and $(-1)^{F_L}$ can also be understood
by going to strong coupling and  using the description by eleven-dimensional
supergravity.  Thus, ${\bf R}^5\times {\bf S}^1\times {\rm K3}$ of the
Type IIA description becomes ${\bf R}^5\times {\bf S}^1\times {\rm K3}\times
{\bf S}^1$, where the radius of the second ${\bf S}^1$ is related
\ewitten\ to the Type IIA string coupling constant.  In this description,
$(-1)^{F_L}$ is the operation that acts as $-1$ on the second circle
and changes the sign of the three-form $A$ of eleven-dimensional supergravity;
$X$ is the operation that acts as $-1$ on the first circle and changes
the sign of $A$.  The two operations are conjugate under exchange of the
two circles.

\bigskip
\noindent{\it Orientifolds}

Now we wish, rather as in \dlp, to discuss orientifolds, by which we mean
orbifolds in which one divides by a symmetry such as $X$ that reverses the
orientation of the world-sheet.  (These generalize the
 Type I superstring, which  should very plausibly
be considered as such an orbifold of the Type IIB theory -- which is invariant
under reversal of world-sheet orientation with no action on space-time.)

First let us describe in general what such an orbifold means.  We consider
a general space-time $M$ with  $\Z_2$ action that reverses the orientation.
The orientifold is then a string propagating on $M/\Z_2$, with a
not-necessarily-orientable world-sheet $\Sigma$, and the following restriction.
One considers only Riemann surfaces $\Sigma$, and maps $\Phi:\Sigma\to
M/\Z_2$, such that the pull-back by $\Phi$ of $w_1(M/\Z_2)$ equals
$w_1(\Sigma)$ (where $w_1$ is the first Steifel-Whitney class,
measuring the obstruction to orientability).  Or more informally,
orientation-reversing loops in $\Sigma$
are mapped to orientation-reversing loops in space-time.  At least when $\Z_2$
acts
freely, it seems  fairly clear that this will give a consistent string theory;
this is presumably true in much greater generality.

We could simply take the orientation reversing symmetry of $\R^5\times
\S^1\times
\K3$ discussed above (acting by $-1$ on $\S^1$ and trivially on the rest)
and attempt to construct the corresponding Type II orientifold.
The fixed points in the symmetry action on $\S^1$ would lead to interesting
subtleties.

If we want a model for which the orientifold can be constructed without such
subtleties, we can compactify further to $\R^4\times \S^1\times \S^1\times
\K3$, and
consider an involution $Y$ that acts by a $\pi$ rotation of the first $\S^1$,
multiplies the second $\S^1$ by $-1$, and reverses the orientation of the
world-sheet.  The orientifold obtained by dividing by this symmetry is a simple
example of a Type II vacuum with $N=2$ space-time supersymmetry realized
by a mixture of left-movers and right-movers.

Moreover, since $Y$ acts freely on $\S^1\times \S^1$, we can -- if the
adiabatic
argument is valid -- determine the heterotic string dual of this model.
We simply consider the heterotic string         on $\R^4\times \S^1\times \S^1
\times \T^4$, and consider the orbifold by the  symmetry $\tilde Y$
that rotates the first circle by $\pi$, and acts as $X$    on $\S^1\times
\T^4$.  (As we noted when $X$ was first defined, the criteria leading to its
definition do not uniquely determine whether $X$ should act by a shift on the
fixed $\Gamma^{1,1}$.  If such a shift was omitted in defining $X$ -- natural
to make $(-1)^{F_L}$ and $X$ precisely conjugate -- then it must be included
here for level matching in  the orbifold.)

So far, we have constructed two elementary examples of Type II vacua
with  $N=2$ space-time supersymmetry: one with $(1,4)$ world-sheet
supersymmetry,
and one an orientifold.  The two models are actually equivalent (after
compactification of the first model from $\R^5$ to $\R^4\times \S^1$) because
 the symmetries $(-1)^{F_L} $ and $X$ are conjugate.
  The equivalence of the orientifold to models
with more familiar world-sheet structures certainly indicates that it
is consistent.

\def\T{{\bf T}}
\def\Z{{\bf Z}}
\subsec{Another $N=2$ Orientifold Example}

In this subsection we will construct another type II $N=2$ orientifold
whose heterotic dual is the same as the one studied in \fhsv .
This orientifold construction has the advantage that our adiabatic
argument applies.  (The model of \fhsv , though well-supported
by other arguments,
cannot be justified by the adiabatic argument because although the orbifolding
group acts freely on the total space, it does not act freely when projected to
the
$\T^2$ that the heterotic and Type II theories share in common.)

Let us first briefly recall the model of \fhsv.
We have on the Type II side
a $\K3\times \T^2$ compactification, for which there is an involution
on $\K3$ giving an  Enriques surface.  We mod out by this $\Z_2$ accompanied
with a total reflection on $\T^2$.  On the heterotic side we first
have to consider the action of $\Z_2$ on the cohomology of $\K3$ .  For
$\K3$'s which double cover the Enriques surface, it is convenient
to write the cohomology lattice as
\eqn\deco{\Gamma^{20,4}=\Gamma^{9,1}_a \oplus \Gamma^{9,1}_b
\oplus \Gamma^{1,1}_a \oplus \Gamma^{1,1}_b}
The ${\bf Z}_2$ action which gives the Enriques surface acts
on the $K3$ cohomology by
\eqn\enri{|\gamma^{9,1}_a,\gamma^{9,1}_b,
\gamma^{1,1}_a,\gamma^{1,1}_b\rangle
\rightarrow |\gamma^{9,1}_b,
\gamma^{9,1}_a,-\gamma^{1,1}_a,\gamma^{1,1}_b\rangle}
The ${\bf Z}_2$ action
$h$ considered in \fhsv\ combines this involution with the
involution on the common $T^2$, as well as some phase factors
\eqn\ffh{
h|\gamma^{9,1}_a,\gamma^{9,1}_b,\gamma^{1,1}_a,\gamma^{1,1}_b,\gamma^{2,2}
\rangle
= {\rm exp}(2\pi i A_2\cdot \gamma^{1,1}_b)|\gamma^{9,1}_b,
\gamma^{9,1}_a,-\gamma^{1,1}_a,\gamma^{1,1}_b ,-\gamma^{2,2}\rangle}
where the $\gamma^{2,2}\in \Gamma^{2,2}$ denotes a vector in
the Narain lattice for the common $\T^2$.
The massless spectrum includes 11 vector multiplets parametrized
by the standard $SO(10,2)\times SL(2)$ coset, and 12  hypermultiplets
parametrized by the standard $SO(12,4)$ coset.  The dilaton on the
heterotic side is part of the $SL(2)$ coset but on the  Type II side is part
of a hypermultiplet.  It was argued in \fhsv\ using
duality that the classical
geometry of neither multiplet receives quantum corrections.

Now we wish to describe the same model
by a Type II orientifold.  The model
is basically the same as the $(1,4)$ model
discussed in the previous subsection,
except that we accompany the orientifold action with the involution
on $\K3$ which gives the Enriques surface.
To describe this in detail, let us introduce some notation.
We consider the common $\T^2$ as a product $\S^1_c\times \S^1_d$.
We consider the orientifold action which acts by a reflection on
$\S^1_c$, translation by $\pi$ on $\S^1_d$, Enriques involution
on $\K3$ and left-right exchange on the world sheet.  Note that to
write the orientifold action on the $\K3$ cohomology in the decomposition
of the lattice given in \deco , the first three factors combine
to give $\Gamma^{19,3}$ and the last one is $\Gamma^{1,1}_b$.
As discussed before the orientifold action flips the sign of one
of these two sublattices.  In this case, unlike  the previous
case, in order for $X^2=1$ the action is reflection on the $\Gamma^{1,1}_b$.
Taking all this into account and noting how the Enriques involution
acts on the $\K3$ cohomology \enri\ we find on the heterotic side
\eqn\hetact{{\tilde h}
|\gamma^{9,1}_a,\gamma^{9,1}_b,\gamma^{1,1}_a,\gamma^{1,1}_b,\gamma^{1,1}_c,
\gamma^{1,1}_d
\rangle
= {\rm exp}(2\pi i A_2\cdot \gamma^{1,1}_d)|\gamma^{9,1}_b,
\gamma^{9,1}_a,-\gamma^{1,1}_a,-\gamma^{1,1}_b ,-\gamma^{1,1}_c
,\gamma^{1,1}_d\rangle}
where just as in the previous example the level matching forces us
to choose $A_2$
instead of $A_3$ on the heterotic side.
Note that $\tilde h$ is conjugate to $h$ given in \ffh\ and
so the resulting heterotic strings are isomorphic.  The main difference
 is that a circle which is part of the cohomology of $\K3$ in the
previous case has been exchanged with a  circle which is common
to Type II and heterotic strings.

Though this orientifold construction has the advantage of satisfying
the adiabatic argument, it has the disadvantage compared
to the construction of \fhsv\ that
we cannot determine using it
whether there would be any quantum corrections,
because as discussed before the dilaton on the type II side is now
a linear combination of
scalars in a hypermultiplet and a vector multiplet.

\subsec{$N=1$ Candidate}

In this section, we will propose candidate $N=1$ dual pairs.
The candidates are obtained by combining the construction of \fhsv\  with
either orbifolding by $(-1)^{F_L}$ or orientifolding.  We give first the
former description.  We
 can also choose the orientifold version of that model discussed
above, giving a construction much like what follows.

We wish to consider a ${\bf Z}_2 \times {\bf Z}_2$ orbifold
generated by $g$ (given in \ofact ) and $h$ (given in \ffh ).
Actually in order to have level matching in the $gh$ sector
for the heterotic string we need to modify the $g$ action of \ofact\
by
\eqn\jikko{g\rightarrow g \cdot {\rm exp}(2\pi i (A^a_2\cdot \gamma^{1,1}_a))}
with $A_2$ as defined before.
The first point of this formula is that (if one does not consider also $h$)
it is equivalent to $g$ as defined in \ofact.  The reason is that $g$ in
\ofact\
acts as $-1$ on $\Gamma_{a}^{1,1}$.  A $-1$ action together with a shift
is equivalent to a $-1$ action, so \jikko\ is equivalent to the earlier
definition.  But this modification of $g$ (or an equivalent modification of
$h$) is needed to construct the ${\bf Z}_2\times {\bf Z}_2$ orbifold.

The second point is that with this modification of $g$,
 the $gh$ twisted sector has the
same structure as the $h$ twisted sector with the role of $\Gamma^{1,1}_a$
and $\Gamma^{1,1}_b$ exchanged.  Note that this modification
does not alter the massless spectrum discussed before for that model.
To count how many supersymmetries we are left with, note that
the action of $g$ and $h$ on the right-moving $T^6$ of heterotic string
in a complex basis is of the form $(-1,-1,1) $ and $(-1,1,-1)$,
which thus gives us $N=1$ spacetime supersymmetry in four dimensions.
It is easy to see that generically we get no massless states
from twisted sectors.  To count how many massless fields we
have it suffices to consider modding out the action of $h$ on the
massless spectrum of $(1,4)$ model discussed above.  One easily
finds that no gauge fields will survive the projection.  At
weak coupling,
the massless moduli that were previously given locally
by the coset of $SO(4,20)
\times SL(2)^3$
are now given by the coset of $SO(2,10)\times SO(2,10)\times SL(2)^3$.
The factors are respectively the complex structure and Kahler moduli spaces
of the Calabi-Yau manifold studied in \fhsv, and the $S$, $T$,
and $U$ moduli spaces.
In
particular, there are $10+10+3=23$ massless $N=1$ chiral multiplets.
We will determine later the details of the weak coupling identifications
of the moduli spaces on the two sides.

For this to make sense, one would hope that the model
in question does not generate a spacetime superpotential and has unbroken
supersymmetry on the whole classical moduli space just discussed.
A preliminary point in favor of this is that as the gauge group is generically
trivial, one does not have gluino condensation generating
a superpotential at the field theory level.

For a further test of a similar nature, we should consider the
 extra massless states that arise  at special points on the moduli space.
If, for instance, at a special point on the moduli space one would find
an $SU(2)$ gauge theory with a pair of doublet chiral superfields
-- a model that in field theory generates a nonperturbative
superpotential  with a pole
at the point where naively the $SU(2)$ is unbroken \ref\ads{I. Affleck,
M. Dine, and N. Seiberg,``Dynamical Supersymmetry Breaking in Four-Dimensions
and its Phenomenological Implications,'' Nucl. Phys. {\bf B256} (1985) 557.}
 -- then one could
very plausibly (though not rigorously) expect    the string theory to generate
a nonperturbative superpotential with such a pole.  However, this
does not occur; on the contrary, in the model considered here, the massless
spectrum at points of  extended gauge symmetry always corresponds to a field
theory that is not asymptotically free.

Extra massless states appear when the Narain lattice corresponds
in the untwisted heterotic string to a point of enhanced gauge symmetry;
 modding out by
$g$ and $h$ then  keeps some of the extra massless particles.  Let us examine
 the enhanced gauge symmetry points
considered in \fhsv\ and see what happens to them after we project
by $g$.  Two cases were considered in \fhsv : Level $2$
ADE type groups coming from projecting to the invariant subsector
of the $h$ action (which exchanges $\Gamma^{9,1}_a\leftrightarrow
\Gamma^{9,1}_b$) with an adjoint hypermultiplet,
 or $SU(2)$ with 4 hypermultiplet doublets
coming from the twisted sectors. These representations lead to finite
(conformally invariant) $N=2$ field theories at low energies in four
dimensions, and have the rather special further
property that in those low energy field theories, there are no quantum
corrections  to the classical metric.
 Projecting these enhanced gauge groups
by $h$ to obtain an $N=1 $ model gives  gauge groups with  matter in a
 representation large enough to spoil asymptotic freedom.
For example for the points with a level two   $SU(n)$ symmetry in the $N=2$
model,
after $g$ projection we get an  $SO(n)$ gauge theory
with matter in one copy of the adjoint representation
of $SO(n)$ and two in the symmetric traceless
tensor product of two fundamental representations.  For an $SO(2n)$ point of
the $N=2$ theory, after $g$ projection
we get $SO(n)\times SO(n)$ gauge group, with one chiral superfield
 in the adjoint representation
and two in the $(n,n)$ representation.  For the point that gives in the $N=2$
theory
 an $SU(2)$
theory with four pairs of doublets,
the gauge symmetry is reduced to $U(1)$ by the $g$ projection, and
some charged matter survives.  In all these cases the spectrum
we get is not asymptotically free.  This means that for
weak coupling the dynamics of the gauge sector at these special
points is  trivial.
These results are compatible with the hypothesis that the theory does not
generate a superpotential.

If this is the case, one may ask how a low energy observer in this theory
interprets
the absence of a superpotential.
The reason that this may come as a surprise is that in renormalizable
field theory in four dimensions, nonperturbative
superpotentials are usually generated unless this is prevented by continuous
global symmetries.  String theory does not have continuous global symmetries,
and discrete global symmetries  could generally not forbid a superpotential.

We will now explain how under suitable conditions,
 a low energy observer who  can probe the moduli
space $M$ of vacua of the theory and is familiar with all the dualities could
understand the
absence
of a string-generated superpotential in a  theory such as this one.
We consider for simplicity in exposition $S$ and $T$ only and
suppose first that the moduli space is $M={\bf P}_S\times {\bf P}_T$, where
${\bf P}_S$ and ${\bf P}_T$ are $S$ and $T$ moduli spaces, compact except
for weak coupling and large volume limits.
If the superpotential is non-zero, it must have a pole somewhere on $M$.\foot{
Thus, in attempts to guess superpotentials invariant under some form of
$S$-duality, poles were required \ref\shapere{S. Ferrara, D. Lust,
A. Shapere and S. Theisen,``Modular Invariance in Supersymmetric
Field Theories,'' Phys. Lett. {\bf B225} (1989) 363.}.}
Since the curves $S=\infty$ and $T=\infty$ (weak coupling and large volume)
generate the Picard group of $M$,
the
pole must intersect either $S=\infty$ or $T=\infty$.  Since $S=\infty$ and
$T=\infty$ are both weak coupling in one theory or the other, the pole
would have to result from a physical phenomenon (such as unbroken $SU(2)$ with
two doublets) that is visible in one description or the other.  In a model
(such as the one that we have discussed) with no such phenomenon,
 there would be  no pole and no superpotential.

\bigskip\noindent{\it Critical Look}

Let us look at that argument more critically.  The assertion that the moduli
space is a product ${\bf P}_S\times {\bf P}_T$ has three characteristics:
 (a) it may not be true even
in weak coupling; (b) it is a stronger statement than we need; (c)
for the statement to be useful, there should be a practical way of verifying
it. We consider the three points in turn.

For (a), we parametrize the weakly coupled theory by the complex variable
$q_S=e^{-S}$.  In string perturbation theory, $|q_S|$ is visible as
the microscopic string coupling
constant and is naturally well-defined independent of $T$ and other moduli.
That is not so for the argument of $q_S$, whose zero mode is the axion,
which is hard to see in string perturbation theory.   Around a divisor
in ${\bf P}_T$ along which charged fields become massless, the axion jumps
by an integer multiple of $2\pi$.  Geometrically this means that even
near $q_S=0$, the moduli space is not a product  ${\bf P}_S\times {\bf P}_T$
but a complex line bundle over ${\bf P}_T$.
This will commonly occur on the heterotic side at enhanced gauge symmetry
points.  Conversely, near the zero locus of $q_T=e^{-T}$, which can be
detected in weak coupling of Type II, one may get a fiber bundle or product
structure because of either conifold singularities or monodromy near $S=
\infty$ where the Type II volume goes to infinity and
the four-dimensional description breaks down.

Since these phenomena will commonly occur, it is fortunate that as asserted
in (b), something much weaker than a product structure will suffice for
our analysis.  It suffices, in fact, to have a {\it global} fiber bundle
structure over either ${\bf P}_S$ or ${\bf P}_T$, together with compactness
of the total moduli space (the compactness modulo weak coupling and large
volume limits by now appears to be a general
property of string theory).  For such a global
fiber bundle, the fiber and base generate the Picard group (one can argue
this by using an adiabatic argument or spectral sequence to compute the
cohomology of the total space), so that any
pole can be seen at weak coupling on one side or the other.
This type of argument still works when additional moduli such as $U$
are included, and it is good enough if the actual moduli space is only
birational to a global fiber bundle.

What about (c), the problem of justifying such a {\it global} fiber bundle
structure?   We will give an example of a simple criterion that is likely
to be useful.  Suppose that in weak coupling
to first order for $q_T$ near zero, one has a product structure
${\bf P}_S\times
{\bf P}_T$,
and that moreover $H^1({\bf P}_S,{\cal O})=H^1({\bf P}_S,T{\bf P}_S)=0$
(${\cal O}$ and $T{\bf P}_S$ are the trivial bundle and tangent bundle
over ${\bf P}_S$).
Then the
desired structure follows.\foot{One proves this by comparing the complex
structure of the moduli space to that of ${\bf P}_S\times
{\bf C}$
(${\bf C}$ being the $q_T$ plane), order
by order in $q_T$, near $q_T=0$.  They must coincide order by order
if ${\bf P}_S\times {\bf C}$ is rigid, that is if its cohomology with
values in the tangent bundle vanishes.  This is so if $H^1({\bf P}_S,
{\cal O})=H^1({\bf P}_S,T{\bf P}_S)=0$.  Once the equivalence is known
to all finite orders, it follows as an exact statement (up to a birational
transformation and finite cover)
by general facts about compact algebraic manifolds.}
Some variants of this argument go through, under certain conditions,
when the structure near $q_T=0$ is a fiber bundle rather than a product.

To see that the criterion just stated is likely to be useful, note that
at the end of this section, we will determine ${\bf P}_S$, for a model
such as this one which is constructed on the Type II side by dividing ${\bf
T}^2$
by a null vector of order two, to be the quotient of the upper half plane
by $\Gamma_0(2)$.  As this quotient is a copy of ${\bf CP}^1$, the relevant
cohomology groups vanish.  So our criterion applies and the space-time
superpotential vanishes if the local structure near ${\bf P}_S$ is a product,
or a line bundle of negative curvature.  This appears plausible, the only
issue, which we will not address here, being a possible monodromy of $q_T$
near $S=\infty$.

\bigskip\noindent
{\it Orientifold Description}

Because of the close relation that we have found between $(-1)^{F_L}$ and
orientifold symmetries, we can also give an orientifold description of the same
model.
We recall the $N=2$ orientifold presented at the end of the last subsection.
We started with the Type IIA theory compactified on
$\R^4\times \S^1\times \S^1\times \K3$, and considered
the $\Z_2$ generated by an operation $Y$ that acts by $\pi $ rotation
of the first circle, multiplication of the second circle by $-1$, and
reversal of world-sheet orientation.  Upon dividing by $\Z_2\times \Z_2$, where
the second $\Z_2$ is the one used in \fhsv\ and described in \ffh,
we get an orientifold with $N=1$ space-time supersymmetry.
By simply translating the description of the various symmetries to the
heterotic
string side, we can describe, as above, a candidate for a heterotic string
dual of this $N=1$ model.  In fact, because of the close relation between
$(-1)^{F_L}$ and orientifolding, the $N=1$ models obtained this way on
the heterotic string side are equivalent -- whether the duality is valid or
not.

\subsec{Asymmetric $N=4$ Models}

We want to determine the weak coupling dualities of the model
above, but first we will describe another interesting model
-- an $N=4$ model studied by Chaudhuri, Hockney, and Lykken \chl\ --
whose dualities can be studied similarly.

According to Chaudhuri and Polchinski \polchinski,  one of the
models in \chl\ in nine dimensions can be constructed as follows.
After compactifying on
${\bf R}^9\times {\bf S}^1$, one takes the orbifold by a certain
operation $f$ that consists of a half-lattice  shift of ${\bf S}^1$
together with an exchange of the two $E_8$ lattices.  To
find a   Type II dual of this model after reduction to
 five dimensions we simply
proceed as follows.
\foot{This has also been analyzed by P. Aspinwall \ref\aspinwall{P. Aspinwall,
lecture at ICTP Workshop on Duality and Mirror Symmetry (June, 1995).}.
  The model actually
has a more challenging  Type II dual in six dimensions
 \ref\schw{J. H. Schwarz and A. Sen,
``The Type IIA Dual Of The Six-Dimensional CHL Compactification,''
hep-th/9507027.}.}
We replace ${\bf R}^9$ by ${\bf R}^5\times
\T^4$, and begin on ${\bf R}^5\times {\bf S}^1\times {\bf T}^4$.
Then we apply string-string duality, considering instead a Type IIA
theory on ${\bf R}^5\times {\bf S}^1\times \K3$.  Now we want
to find the analog on the Type IIA side of dividing by $f$.
Since $f $ acts freely on ${\bf S}^1$, the adiabatic argument
applies and if we can identify the operation $f'$ that corresponds
to $f$,  the
heterotic string orbifold of ${\bf R}^5\times {\bf S}^1\times \T^4$
by $f$ should be equivalent to the Type IIA orbifold of
${\bf R}^5\times {\bf S}^1\times \K3$ by $f'$.

What is $f'$?  Obviously, $f'$ acts by a $\pi$ rotation of
${\bf S}^1$, but how does it act on $\K3$?  It must act
in such a way as to exchange the two $E_8$ lattices in the
cohomology of $\K3$ while fixing the rest.

To find an explicit $\K3$ automorphism that does that, it is helpful
to first understand explicitly what kind of $\K3$ can exhibit
the $E_8\times E_8$ gauge symmetry that is familiar in the heterotic
string.  For this, according to section 4.6 of \ewitten\ the $\K3$ must have
two $E_8$ singularities.
We recall that an $E_8$ singularity is the singularity described
by the equation
\eqn\ppol{w^2+x^3+y^5 = 0}
in ${\bf C}^3$.  To find a $\K3$ with   two $E_8$ singularities,
we simply note that a $\K3$ can be described by the equation
$w^2=P_6(x,y,z)$ where $P_6$ is a generic homogeneous sixth
order polynomial ($x,y,z$, and $w$ are homogeneous coordinates
of weights $1,1,1,3$ in a weighted projective space).  If we pick
a particular $P_6$ giving the  equation
\eqn\opol{w^2+(x-z)^3(x+z)^3+zy^5=0,}
we get a $\K3$  that has two $E_8$ singularities -- at
$(x,y,z,w)=(\pm 1, 0,1,0)$ -- and is otherwise non-singular.
So the Type IIA theory compactified on this $\K3$ can have
$E_8\times E_8$ gauge symmetry.

Now consider the automorphism $h(x,y,z,w)=(-x,y,z,-w)$ of this
$\K3$.  This transformation
exchanges the two $E_8$ singularities,
but  preserves the orientation, complex structure, and
holomorphic two-form, and can be seen to leave invariant the
polynomial deformations that keep two $E_8$ singularities.  So it
exchanges the two $E_8$ lattices and leaves fixed the rest of the
cohomology.
The operation   $f'$ that we want is therefore simply $h$ combined
with a $\pi $ rotation of the circle.

It was noted by Chaudhuri and Polchinski that this  model
seems to exhibit $S$-duality when further compactified to
four dimensions.  Now we can see why that must be so.
The four-dimensional model is an orbifold of
$\R^4\times \S^1\times S^1\times \T^4$; its Type II dual
is an orbifold of $\R^4\times\S^1\times \S^1\times \K3$.
The Type II model has a $T$-duality acting on $\S^1\times \S^1$,
and by a standard argument, this implies  $S$-duality for
the heterotic version of the theory (and vice-versa).  The duality
groups are proper subgroups of $SL(2,{\bf Z})$ that will be analyzed
presently.

\subsec{Weak Coupling Dualities}

\def\K3{{\rm K3}}
\def\Z{{\bf Z}}
\def\T{{\bf T}}
\def\CP{{\bf CP}}

If one compactifies ten-dimensional string theory to four dimensions
on ${\bf T}^2\times X$, with $X$ any four-manifold,
one gets a $T$-duality group $SL(2,
\Z)_T\times SL(2,\Z)_U$; string-string duality exchanges $S$ and $T$,
adding a third copy of $SL(2,\Z)$.
We want to consider the case in which $\T^2$ has been divided by a $\Z_2$
shift of the Narain lattice, perhaps acting also on other degrees of freedom
and determine the weak coupling duality group.  We first make a general
discussion and then apply it to several models constructed above.

First we describe the Narain lattice in a way that exhibits the $SL(2)\times
SL(2)$ symmetry.  We consider $2\times 2$ integer matrices
$M^{ii'},\, i,i'=1,2$, with inner product $(M,N)=\epsilon_{ij}
\epsilon_{i'j'}M^{ii'}M^{jj'}$ (so for instance $(M,M)=2\det M$).
$SL(2)\times SL(2)$ acts by $M\to AMB^{-1}$ with $(A,B)\in SL(2)\times SL(2)$.
There are $2^4-1=15$ non-zero half-lattice vectors, modulo lattice shifts.
Write such a vector as $V=M/2$, with $M$ a lattice vector.  One can think
of $M$ as a non-zero matrix whose entries are all $0$ or $1$.  Nine such
matrices have determinant zero, giving $V^2=0$.  It is easy to check that
such matrices form an irreducible orbit of $SL(2)\times SL(2)$.   A
representative such matrix is
\eqn\jjhj{\left(\matrix{1 & 0 \cr 0 & 0 \cr}\right)}
and is obviously stabilized by a subgroup conjugate to $\Gamma_0(2)
\times\Gamma_0(2)$ ($\Gamma_0(2)$ is the index three subgroup of $SL(2,\Z)$
consisting of matrices that stabilize the vector $(1,0)$ modulo two).
As $\Gamma_0(2)\times \Gamma_0(2)$ is of index nine in $SL(2,\Z)\times
SL(2,\Z)$, it is the full stabilizer of the half-lattice vector determined
by \jjhj, and hence is the weak coupling $T-U$ duality group of a theory
constructed by orbifolding by such a shift.

The other case is a matrix of determinant $\pm 1$, such as $M=1$.  It is again
easy to check that the six such matrices are permuted by $SL(2,{\bf Z})\times
SL(2,{\bf Z})$.  The stabilizer of $M=1$ consists of pairs $(A,B)$ with
$AMB^{-1}=M$ modulo two, that is, $A=B$ modulo two.  Thus, the $T-U$
duality group of a model constructed by orbifolding by such a shift
is the subgroup $H$ of $SL(2,\Z)\times SL(2,\Z)$ consisting of pairs
$(A,B)$ with $A=B$ modulo two.  This subgroup is of index six, as expected.

Now we can determine the duality group of the model of the CHL model
 compactified to four dimensions.  This model is constructed
on both heterotic and Type II sides by dividing by a null shift vector,
so the $S$, $T$, and $U$ duality groups are all $\Gamma_0(2)$.  This shows,
in particular, that just because a model has $N=4$ supersymmetry in four
dimensions does not mean that the $S$-duality group will turn out to be all
of $SL(2,{\bf Z})$.  Similarly, many different models  with the same
low energy physics are likely to have different duality groups at the string
level.  For instance, $N=5$ and $N=6$ supersymmetry, which can be constructed
in many ways using asymmetric Type II orbifolds, give low energy
duality groups with no canonical integral structure, so that different
string theory realizations very likely have different duality groups.

Finally, let us consider the $(1,4)$ model with $N=2$ space-time supersymmetry
that was constructed above in several ways.  On the heterotic string
side, this is constructed with a shift of non-zero length squared,
so the heterotic string weak coupling
$T-U$ duality group is the group $H$ described above.  On the Type II
side, the shift is by a null vector, so (remembering the exchange of $T$ and
$S$), the $S-U$ duality group at Type II weak coupling is $\Gamma_0(2)\times
\Gamma_0(2)$.  Note that the two determinations of the $U$-duality group
appear to disagree.  We interpret this as further evidence
 that for $N=2$ there are
corrections to the whole vector moduli space.
For $N=4$ such correction are not expected.

\newsec{$\K3$ Fibrations And String-String Duality}

In \kachruvafa, various examples were constructed
of apparent dual pairs with $N=2$ supersymmetry in four dimensions.
The examples involved, on the heterotic string side,
compactification more or less on $\T^2\times \K3$ with
some vector bundle (in some cases one uses an enhanced gauge
symmetry that comes upon taking the $\T^2$ at a special radius),
while on the Type II side they involved compactification on
certain Calabi-Yau manifolds.  It has been noted \lerche\
that the Calabi-Yau manifolds that arose in these examples
have the structure of $\K3$ fibrations, that is, fiber bundles
over ${\bf CP}^1$ with $\K3$ fibers.

The question arises of whether the structure of $\K3$ fibrations
can actually be used to explain the examples of \kachruvafa,
via the adiabatic argument given in the last section.  Going
to a region in parameter space in which the ${\bf CP}^1$ has a
very large area, while the volume of the $\K3$ fiber is of
order one, one has a Calabi-Yau manifold $X$ which is a family
of slowly varying $\K3$'s.  In the limit that the family varies
very slowly, it should be possible to apply string-string duality
fiber-wise, replacing the Type II description by a heterotic string
description and replacing the $\K3$ fibers by $\T^4$ fibers.
Thus the heterotic string description should involve a family
of $\T^4$'s fibered over ${\bf CP}^1$.  The   $\T^4$'s, along with
the abelian gauge fields that arise at a generic point in
moduli space, are really described by
a Narain lattice $\Gamma^{20,4}$; the variation of this
Narain lattice over ${\bf CP}^1$  should be determined by the
variation of the original family of $\K3$'s.

To test this interpretation in
a preliminary way, note that the $\T^2\times \K3$
used in \kachruvafa\ on the heterotic string side can very well
arise as a family of $\T^4$'s fibered over $\CP^1$.  In
fact, $\K3$ can be realized as a family of $\T^2$'s fibered
over $\CP^1$ (giving an ``elliptic surface''--
see \ref\stringy{B. Greene, A. Shapere, C. Vafa and S.-T. Yau,``
Stringy Cosmic Strings and Noncompact Calabi-Yau Manifolds'',
Nucl. Phys. {\bf B337} (1990) 1.} for the construction
of a  Ricci-flat metric on this fibered space), so $\T^2\times \K3$
can certainly arise as a family of $\T^4$'s fibered over $\K3$.

The main obstruction to this attempt at explaining the results
of \kachruvafa\ is that there are certain singularities in
the $\K3$ fibers, so that the adiabatic approximation is not
everywhere valid.  In this section, we will examine the $\K3$
fibrations that appear in these particular examples, and show
that they have a particularly  simple structure which further
simplifies in the weak coupling limit (of the heterotic string).
The singularities at which the adiabatic argument fails
are of a very special nature.
Nonetheless, not understanding them,
we will not be able to explain the results
of \kachruvafa.  We will, however, be able to deduce from
the structure of $\K3$ fibrations a simple ``orientifolding''
operation which -- if the adiabatic argument is valid -- should
give for every $N=2$ dual pair of this type an $N=1$ dual
pair (or in general several of them).

\subsec{The $\K3$ Fibrations}

String-string duality naturally maps the heterotic string to
a Type IIA string.  If one then uses mirror symmetry to convert
to a Type IIB description, then the vector moduli space of the
heterotic string is mapped to the complex   structure moduli
space of a Type IIB model, which has no world-sheet or
space-time corrections and can be conveniently  studied classically.

We will study two Type II examples from \kachruvafa\ for which the
evidence for a heterotic string description was particularly
strong.
The first example, which we will call $X$, is defined by a
hypersurface  of degree twelve in a weighted
projective space with weights $1,1,2,2,6$:
\eqn\gurf{z_1^{12}+z_2^{12}+z_3^6+z_4^6+z_5^2
-12\psi z_1z_2z_3z_4z_5 -2\phi z_1^6z_2^6=0.}
We recall that a weighted projective space is defined by
identifying the variables under
\eqn\jurf{z_i\to t^{d_i}z_i}
with $d_i$ the weights and $t\in   {\bf C}^*$.
The second example, which we will call $Y$, is defined
by a hypersurface of degree 24 in a weighted projective space with
weights $1,1,2,8,12$:
\eqn\urf{z_1^{24}+z_2^{24}+z_3^{12}+z_4^3+z_5^{2}
-12\alpha z_1z_2z_3z_4z_5 -2\beta z_1^6z_2^6z_3^6-\gamma z_1^{12}
z_2^{12} = 0.}
Actually, in each case (for Type IIB) one is really interested in certain
orbifolds of these spaces, dividing by the group of diagonal
phase rotations of the variables that preserve
the hypersurfaces and the holomorphic three-forms.  The polynomials
written are the most general ones of the correct degrees compatible
with that symmetry (and these particular models have no
non-polynomial deformations).

To see the structure of $\K3$ fibration, we follow the first
few pages of \ref\morrison{P. Candelas, X. De la Ossa, A. Font,
S. Katz and D. Morrison,``Mirror Symmetry for Two Parameter
Models-I,'', Nucl. Phys. {\bf B416} (1994) 481.}.  We consider the submanifold
$H_\lambda$ of $X$ defined
by $ z_1=\lambda z_2$ with some fixed complex $\lambda$.
Once $z_1$ is eliminated in this way, $z_2$ is the only
remaining variable of odd weight, and therefore if we set
$t=-1$ in \jurf, we get the transformation $z_2\to -z_2$, with
other variables invariant.  We can divide out this equivalence
by setting $y=z_2^2$, and after doing this, the equation
for $H_\lambda$ becomes
\eqn\gurfo{(1+\lambda^{12}-2\phi\lambda^6)y^6+z_3^6+z_4^6+z_5^2
-12\lambda \psi yz_3z_4z_5 =0.}
We recognize this as an equation describing a $\K3$ manifold
in a weighted ${\bf  CP}^3$.  So the $H_\lambda$'s are copies
of $\K3$, and $X$  can be regarded as a family of $\K3$'s
parametrized by  $\lambda$, that is, $X$ is
fibered over ${\bf CP}^1$ (the Riemann sphere built from the
$\lambda$ plane) with the fibers being $\K3$'s.  Similarly,
upon setting $z_1=\lambda z_2$ and $y=z_2^2$, \urf\ becomes
\eqn\urfo{(1+\lambda^{24}-\gamma\lambda^{12})y^{12}
 +z_3^{12}+z_4^3+z_5^{2}
-12\alpha \lambda yz_3z_4z_5 -2\beta \lambda^6y^6z_3^6 = 0}
which again defines a $\K3$ manifold, so we again get a family
of $\K3$'s fibered over ${\bf CP}^1$.

Since our plan is to carry out string-string duality
fiber-wise, we need to understand the monodromy of these families
of $\K3$'s.  Since ${\bf CP}^1$ is simply-connected,
such monodromy is only possible because at some values of
$\lambda$ the $\K3$'s develop singularities; the singularities
are the reason that fiber-wise application of string-string
duality may have difficulties.  We have to determine
the monodromy representation and translate it to the heterotic
string side.  One might worry that we will meet very complicated
$SO(20,4;{\bf Z})$-valued monodromies, but this is not so.
The monodromy representations for the above families
of $\K3$'s are as simple as one could hope for.

To see this, set $\psi=0$ in \gurfo\ (or $\alpha=\beta = 0$ in
\urfo).  Then one finds that, for generic $\lambda$,
the $\K3$'s in \gurfo\ and \urfo\ have a structure that is
{\it independent of} $\lambda$, since $\lambda$ can be eliminated
by rescaling $y$.  This fails (and the $\K3$ is singular)
precisely at zeroes of the function $F(\lambda)=1+\lambda^{12}
-2\phi\lambda^6$ (or $G(\lambda) = 1+\lambda^{24}-\gamma
\lambda^{12}$ in the second example).  The monodromy of
the $\K3$ around  zeroes of $F$ or $G$ is very simple.
Near a value of $\lambda$ at which $F$ or $G$ has a zero of
order $n$, the monodromy is given by the orbifold operation
$y\to \zeta^ny$, where      $\zeta=e^{2\pi i/6}$ in the first
example, and $\zeta=e^{2\pi i/12}$ in the second example.

Let us compute how $S:y\to \zeta y$ acts on the cohomology
of these $K3$'s.  Obviously, $H^{0,0}$ and $H^{2,2}$ are
invariant.  $H^{2,0}$ is generated by the holomorphic two-form
$\omega= (y \, dz_3\,dz_4+\dots)/(\partial P/\partial z_5)$,
\foot{In this and similar formulas below, the $\dots$ refers to a
addition of terms obtained by cyclic permutation of the variables; in the
particular case here, $y\,dz_3\,dz_4+\dots = y\,dz_3\,dz_4+z_3\,dz_4\,dy+
z_4\,dy\,dz_3$.  $P$ is the defining polynomial of the hypersurface.}
so $S$ acts
as $\zeta $ on $H^{2,0}$ and as $\zeta^{-1}$ on $H^{0,2}$.
Now consider the action on $H^{1,1}$.  The Kahler form
is obviously invariant under $S$.  The action of $S$ on the
rest of $H^{1,1}$ can be conveniently found by considering
$H^1(T)$, the space of deformations of complex structure.
Given a deformation of complex structure represented by a
$T$-valued $(0,1)$ form $\alpha_{\bar j}{}^i$, one maps
to a $(1,1)$ form by multiplying by $\omega$:
$\beta_{i\bar i}=\omega_{ij}\alpha_{\bar j}{}^j$.  Therefore
the transformation law of $\beta$ has an extra $\zeta $ relative
to that of $\alpha$.

Since $H^{1,1}$ is twenty-dimensional for a $\K3$ surface,
and we already know about the Kahler form, we are looking
for 19 more modes.  For the first example, these come from
19 polynomial deformations in $H^1(T)$: modulo the derivatives
of the defining equation of the hypersurface, there are three
polynomials of the form $y^4P_2(z_3,z_4)$, four of the form
$y^3P_3(z_3,z_4)$, five of the form $y^2P_4(z_3,z_4)$,
four of the form $yP_5(z_3,z_4)$, and three of the form
$P_6(z_3,z_4)$.  ($P_k$ is a homogeneous polynomial of degree given by
the subscript.)  Since $y^nP_{6-n}$  transforms under $S$
as $\zeta^n$, the corresponding $(1,1)$ form transforms
as $\zeta^{n+1}$.  Hence for these 19 modes, $\zeta$ and $\zeta^5$
appear three times, $\zeta^2$ and $\zeta^4$ appear four times,
and $\zeta^5$ appears five times.

Now consider the second example.  In this case there are only
18 polynomial deformations, corresponding to the monomials
$y^nz_3{}^{12-n},\,\,\,2\leq n\leq 10$, and  $y^nz_3{}^{8-n}z_4,
\,\,\,\, 0\leq n\leq 8$.  $S$ acts on the corresponding
$(1,1)$ forms as $\zeta^n$ with $3\,\leq n\leq 11$ for the
first series and $1\leq n\leq 9$ for the second series.
One more mode is needed; where is it?
In fact, the $\K3$ defined in \urfo\ has an $A_1$ singularity
at $y=z_3=0$.  There is one collapsed two-cycle sitting
in this singularity, and it is invariant under $S$.

We would now like to see what the heterotic string dual of this
model should look like, assuming that it can be constructed
by applying string-string duality fiber-wise.  Fiber-wise
application of string-string duality will give a family
of $\T^4$'s parametrized by ${\bf CP}^1$; the $\T^4$'s, with
their Wilson lines, are described by a family of Narain
lattices.  Right-movers
and left-movers of the heterotic string come respectively
from the self-dual and anti-self-dual part of the cohomology of
$\K3$.   For the simplest statement, we look at the  invariant
part of the Narain lattice, which comes from the
monodromy-invariant part of the cohomology of $\K3$.

We consider first the second example $Y$.  The monodromy-invariant
part of the cohomology is four-dimensional, spanned by
$H^{0,0}$, $H^{2,2}$, the Kahler class and the class associated
with the $A_1$ singularity.  The signature of this
four-dimensional invariant subspace is $(2,2)$: the self-dual
part is the Kahler class and a linear combination of
$H^{0,0}$ and $H^{2,2}$.  Modulo    questions of torsion,
this suggests that the Narain lattice has an invariant
sublattice $\Gamma^{2,2}$.  This would correspond to a very
special family of $\T^4$'s over ${\bf CP}^1$, consisting
of the sum of a {\it     fixed} $\T^2$, without Wilson lines,
represented by the fixed $\Gamma^{2,2}$, and a variable
$\T^2$, with Wilson lines, represented by a variable
$\Gamma^{18,2}$.  This is exactly the structure proposed
for this example in \kachruvafa:  the heterotic string dual
of the second example was proposed to be a product
$\K3\times \T^2$, with the gauge bundle living entirely on
the $\K3$.   The $\K3$ can take the form of a $\T^2$ bundle
over ${\bf CP}^1$, so this $\K3\times \T^2$ has the structure
of a $\T^2\times \T^2$ bundle over ${\bf CP}^1$ in which
only the first factor varies.

Now let us consider the first example, $X$.  In this case,
the invariant part of the cohomology is only three-dimensional;
we can only split off a lattice of signature $(1,2)$, with
one left-moving and two right-moving moduli.  This
again has a natural interpretation in relation to \kachruvafa,
according to which
$X$ corresponds again to a heterotic string on $\K3\times \T^2$,
but this time the           $\T^2$ is fixed at a point in
moduli space at which one of the left-moving $U(1)$'s is
extended to $SU(2)$, and that $SU(2)$ is then  broken in
the construction of the gauge bundle.  So
the $\T^2$ carries only one free left-moving mode and two
free right-moving modes, in agreement with what we found
for the monodromy-invariant part of the cohomology of $X $.

Our discussion has helped to give a more vivid picture of
how the heterotic string description of \kachruvafa\ is
related to the classical geometry of $X$ or $Y$ via string-string
duality.  Our discussion, however, can be interpreted purely
in classical terms as a computation of the cohomology of $X$
or $Y$ in terms of the monodromy action on the cohomology
of the fiber, and the  cohomology of $X$ and $Y$ were already
matched in \kachruvafa\ with the moduli of the heterotic string
vacua.  So it is not clear that the agreement we found
between the two sides is really independent evidence for
string-string duality.

\subsec{The Metric, Level Matching, and Weak Coupling}

Next let us figure out what the metric of $X $ and $Y$ look
like in the adiabatic limit in which the area of the $\lambda$
plane is scaled up, keeping fixed the restriction of the Kahler
class to the fiber.  The discussion and result
are quite similar to the discussion of stringy cosmic strings
in \stringy .

In the case of $X$, for instance (the other example is quite
similar for the present purposes), let us consider
the structure near an isolated simple zero
of the function $F(\lambda)$,
which we take to be at $\lambda=0$.
The equation for $X$ looks near $\lambda=0$ like $p=0$ with
\eqn\omigo{p=\lambda y^6 +z_3{}^6+z_4{}^6+z_5{}^2.}
In the adiabatic limit, with the metric on the $\lambda$
plane scaled up, the form \omigo\ for the equation becomes
valid throughout the $\lambda$ plane.
Recalling that the holomorphic three-form in the original
variables was $\omega=
(z_1\,dz_2\,dz_3\,dz_4\pm ...)/(\partial p/\partial z_5)$,
we see that in this  description the holomorphic three-form
looks like
\eqn\kkk{\omega=d \lambda\cdot {y\, dz_2\,dz_3\pm \dots \over
\partial p/\partial z_5}.}
One can eliminate $\lambda$ from \omigo\ by setting $\tilde
y =\lambda ^{1/6}y$, whereupon
$\omega$ can be written $\omega=(d\lambda/\lambda^{1/6})\cdot
\Omega$, where
\eqn\dnmdnm{\Omega={\tilde y \,dz_3\,dz_4+\dots\over \partial p/\partial
z_5}}
is a fixed two-form on a fixed $\K3$.

If then one has a Ricci-flat Kahler metric on the total
space of \omigo\ which at large $\lambda$ looks like the product
of a metric on the $\lambda$ plane with a fixed metric
on $\K3$, then the metric on the $\lambda$ plane
will be $ds^2=|d\lambda/\lambda^{1/6}|^2$.  This is a flat
metric (away from the origin) with a deficit angle at infinity
of $2\pi/6$.  Since the polynomial $F(\lambda)$ is
of twelfth order, there are          twelve of these
singularities.  With twelve singularities, the total
deficit angle is $4\pi$.  The interpretation is        clear.
According to the Gauss-Bonnet theorem, the integrated curvature
of any metric on the $\lambda$ plane is $4\pi$.  In the
adiabatic limit, the $\lambda$ plane looks flat except for
twelve delta functions, of strength $2\pi/6$, at zeroes
of $f$; each  delta function arises at a conical singularity
with a deficit angle of $2\pi/6$.

\bigskip
\noindent{\it Orbifold and Level Matching}

It may appear that we can now test string-string duality
in the following way.  In the special case in which
five zeroes of $\lambda$ coincide (so $\lambda$ in \omigo\
is replaced by $\lambda^5$), the deficit angle becomes
$2\pi \cdot 5/6$, which is what we would get if we divide
the $\lambda$ plane by $\lambda\to \zeta \lambda$. So our
Calabi-Yau becomes a $\Z_6$ orbifold of ${\bf C}\times \K3$, with
${\bf C}$ the complex plane
and  $\bf Z_6$ acting by $\lambda\to \zeta\lambda$,
$y\to\zeta^5 y=\zeta^{-1} y$.
It would appear that this should correspond
under string-string duality to a heterotic string
orbifold of ${\bf C}\times {\bf T}^2$ by ${\bf Z}_6$.

To see if this makes sense, we look at the action of
${\bf Z}_6$ on the cohomology of $\K3$, which will determine
the behavior of the left-movers of the heterotic string.
The left-movers with non-trivial monodromy come from modes
of $H^{1,1}$ derived from polynomials and from the $\lambda$
plane.  The eigenvalue $\zeta^s$ arises for left-moving bosons
with multiplicity four for $s=1,2,4,5$ and five for $s=3$.
The left-moving
ground state energy is thus shifted by $\sum_i s_i(6-s_i)/4\cdot 6^2
= 149/144$.  This is not a multiple of $1/6$, so level
matching does not hold.  Moreover, the discrepancy cannot
be corrected by combining the ${\bf Z}_6$ shift
with a shift by $1/6$ of a Narain lattice vector, which
adds to the discrepancy in level-matching  a multiple
of $1/6^2$.

So we get another example of the fact that orbifolding does
not commute with string-string duality: the Type IIA orbifold
of ${\bf C}\times \K3$ by ${\bf Z}_6$ does not  correspond
to a heterotic string orbifold of ${\bf C}\times \T^4$.
Moreover, it might appear that we face a near-paradox
in reconciling the duality proposed in \kachruvafa\ with
what we have learned from the adiabatic argument.

\bigskip\noindent
{\it Weak Coupling}

The paradox can be overcome if we note that heterotic string
level matching is a weak coupling concept so that we must
compare heterotic string level matching conditions to the monodromies
that arise where the heterotic string is weakly coupled.

According to \kachruvafa, the region of moduli space of $X$ in
which the heterotic string is weakly coupled is $\phi\to\infty$.
In that limit, of the twelve zeroes of $F$, six are at the origin
and six are at infinity.  With suitable rescaling of $\lambda$,
$F$ reduces in the limit to $F(\lambda)=\lambda^6$.  The mapping
from the      Type IIA model to the heterotic string is such
that the splittings among the groups of six zeroes are
non-perturbatively small from the standpoint of the heterotic
string.  Thus, in comparing to heterotic string perturbation
theory, we have simply two groups of six zeroes.

For a group of six zeroes, the monodromy disappears!  Our
problem with level matching disappears with it.

We can also now see something about what the metric looks
like in the limit in which the heterotic string is weakly coupled.
Since each       zero of $F$ produces a deficit angle of $2\pi/6$,
when six of them coincide one gets a deficit angle of $2\pi$.  What
a deficit angle of $2\pi$ means is that in the limit that the heterotic
string is weakly coupled, the ${\bf CP}^1$ degenerates to
a semi-infinite cigar with the six zeroes of $F$ coalescing at
the  tip of the cigar and the other end being an infinite      flat
cylinder.  When the heterotic string coupling is very small but
not zero, one has a very long but not infinite cigar with six
zeroes of $F$ at one end and six at the other.

Presumably, in perturbation theory of the heterotic string, one
sees one end of the cigar or the other, but not both.  World-sheet
instantons of the Type II theory, in which the  world-sheet wraps
all the way around the cigar, would see both ends of the cigar.
In the mapping between the Type II and heterotic string descriptions,
these correspond to space-time instantons of the heterotic string.
It would be quite interesting to construct for each such world-sheet instanton
of type IIA,
a spacetime instanton of the heterotic string, ideally in the form
of a $(0,4)$ superconformal field theory.

\bigskip\noindent
{\it A Note On The Example of \fhsv }

We have been discussing the examples of \kachruvafa\ as $\K3$ fibrations.
It is interesting to note that the example of \fhsv\ also has the
structure of a K3 fibration over ${\bf CP}^1$.  This example
was constructed by starting with $\K3\times \T^2$ and acting
by a $\Z_2$ that acts on both factors.  By forgetting the first
factor, $(\K3\times \T^2)/\Z_2$ maps to $\T^2/\Z_2={\bf CP}^1$,
and the fibers are $\K3$'s.  In this case, the monodromies of
the Narain lattice come from the $\Z_2$ action on $\K3$, and in this
particular case, there is no difficulty with level matching in the
weakly coupled limit of the heterotic string.

\subsec{$N=1$ Orientifolds}

One of the most interesting consequences of the structure of
$\K3$ fibrations of the examples considered in \kachruvafa\
is that this makes it possible to identify $N=1$ orientifolds
for which a heterotic string dual can be found in a way that should
be reliable.

First of all, given a Calabi-Yau compactification of the Type IIA
superstring, to construct an orientifold we need to divide by
an isometry that reverses the orientation.  The most obvious
orientation-reversing isometries of Calabi-Yau threefolds are
symmetries that reverse the complex structure.

To begin with, note that ${\bf CP}^1$ has an antiholomorphic
symmetry that acts without fixed points.  In terms of homogeneous
coordinates $(u,v)$, the transformation is $\tau(u,v)= (\bar v,-\bar u)$.
(The quotient of ${\bf CP}^1$ by this transformation is
${\bf RP}^2$.) Suppose that we find an antiholomorphic involution $w$
of one of the above  examples that         preserves the structure
of $\K3$ fibration and acts as $\tau$ on the base.  Then --  upon
taking the area of the base to be very large -- the adiabatic argument
should apply and it should be possible to find a heterotic
string dual by fiber-wise application of string-string duality.

Actually, all of these models have antiholomorphic involutions with
the right properties.  For example, consider the $X$ manifold, defined
by the equation
\eqn\gurfok{z_1^{12}+z_2^{12}+z_3^6+z_4^6+z_5^2
-12\psi z_1z_2z_3z_4z_5 -2\phi z_1^6z_2^6+\dots =0.}
For constructing an  orientifold and applying string-string duality,
it is most natural to consider the Type IIA rather than Type IIB
superstring, so we are interested in the $X$ manifold rather than an
orbifold of it, and many terms can be added to \gurfok.
This manifold
has the antiholomorphic symmetry
\eqn\immo{w(z_1,z_2,z_3,z_4,z_5) = (\bar z_2,-\bar z_1, \bar z_3,\bar
z_4,\bar z_5)}
provided that $\phi$ is real and $\psi$ imaginary (and similar
conditions are put on the other complex parameters).  This acts
as $\tau$ on the base since it reduces to $\tau$ if one forgets
$z_3,z_4$, and $z_5$.   There also
are some variants of this; one can exchange $z_3$ and $z_4$, with
or without a change of sign.

Let us determine the massless spectrum of the orientifold obtained
by dividing by   $w$.  An important question is how     $w$ acts
on $H^{1,1}$.  In this particular example, $w$ acts as $-1$     on
$H^{1,1}$.\foot{$w$ acts as $+1$ on $H^{2,2}$ and therefore, by
Poincar\'e duality (given that $w$ reverses the orientation) it acts
as $-1$ on $H^{1,1}$.  To see that $w$ acts as $+1$ on $H^{2,2}$,
note that $H^{2,2}$ is generated according to \morrison\ by the
divisor $z_1=0$ (which can be deformed into its image under $w$)
and the $w$-invariant divisor obtained by blowing up the curve
$z_1=z_2=0$.}  Now, each mode $\beta$ in $H^{1,1}$ determines in the Type
IIA theory a vector multiplet consisting of a Kahler mode, a mode
of the $B$ field, and an abelian gauge field.
The Kahler mode is made          by multipling $\beta $ by the
complex structure, which is odd under $w$, so the Kahler mode
is even and survives in the orientifold.   The $B$ field mode is
odd under $w$, but also odd under exchanging left and right-movers
on the world-sheet, so it also survives in the orientifold.
As for the gauge field in the vector multiplet, it is projected
out in forming the orientifold.  So $N=2$ vector multiplets
associated with modes in $H^{1,1}$ odd under $w$     are truncated
to $N=1$ chiral multiplets in the orientifold.

In a more general model of this type, $H^{1,1}$ might contain
a subspace even under $w$.  From the corresponding scalar multiplets,
the Kahler deformation and $B$ field would be projected out,
but the gauge field would survive.  So $N=2$ vector multiplets
associated with modes in $H^{1,1}$ even under $w$ are truncated
to $N=1$ vector multiplets in the orientifold.

A hypermultiplet, on the other hand, contains  four states,
two from the NS-NS sector and two from the Ramond-Ramond sector.
In the orientifold, one NS-NS state and one Ramond-Ramond state
survive, making an $N=1$ chiral multiplet.
For example, the fact that \gurfok\ is $w$-invariant
only for $\phi$ real and $\psi$ imaginary means that of the
complex scalars NS-NS scalars associated with $ \phi$ and $\psi$,
only one real component survives in the orientifold.  Its $N=1$
chiral partner survives in the orientifold.

\subsec{Concrete Description On Heterotic String Side}

Now we will describe in a specific case what the heterotic string
equivalent to one of these orientifolds looks like.  We consider on the
Type II side the $Y$ manifold.  We want to translate the orientifold symmetry
of the $Y$ manifold to the heterotic string side and then divide by it.

On the heterotic string side, according to \kachruvafa, the $Y$ manifold
corresponds to compactification on $\T^2\times \K3$, with the following
gauge bundle.  One takes an $E_8\times E_8$ bundle over $\K3$ with second
Chern class twelve on each factor.  The gauge bundle on $\T^2\times \K3$
is thus derived purely from a bundle on $\K3$.  The model has four
vectors coming from the moduli of $\T^2$.

Incidentally, the alternative
case of an $SO(32)$ bundle over $\T^2\times \K3$ which is derived from
a bundle over $\K3$ with second Chern class 24 -- an example also
mentioned in \kachruvafa\ -- can be identified in the following way with
one of the $\K3$ fibrations listed in \lerche.  With standard embedding
of the gauge bundle in the tangent bundle, this model has  gauge group $SO(28)
\times SU(2)$ with ten massless hypermultiplets in the $({\bf 28},{\bf 2})$.
Expectation values of these generically break $SO(28)$ to $SO(8)$ (the complete
Higgsing assumed in \kachruvafa\ to relate this model to the $Y$ manifold
cannot occur).  Going to the Coulomb phase of the $SO(8)$ gives a model
with eight vectors (four from $\T^2$ and four from the Cartan subalgebra
of $SO(8)$) and 272 hypermultiplets, corresponding quite likely
to a $\K3$ fibration
with $h^{1,1}=7,\,\,\,h^{2,1}=271$ that is listed in \lerche.   Similarly,
the heterotic string model in section (4.2) of \kachruvafa\ can
be taken with gauge group $E_8\times E_8\times SU(2)$ or $SO(32)\times SU(2)$.
In the former case, after complete Higgsing one gets the $X$ manifold
with three vectors, as described in \kachruvafa, while in the latter
case the hypermultiplets can break the gauge symmetry only to $SO(12)$,
and one gets
presumably the model with $h^{1,1}=8,\,\,h^{2,1}=194$ given in \lerche.
One should aim to similarly reproduce from the heterotic string
many examples in \lerche, and their
toric generalizations.

The heterotic string analog of the orientifold symmetry  $w$ of
the $Y$ manifold
can be studied as follows.  First of all, $w$ must act as $-1$ on
$\T^2$ so that the four vectors coming from $\T^2$ are all odd under $\tau$.
To find the action on $\K3$, we use the fact that the $\K3$ in question
is fibered over ${\bf CP}^1$ and that the $w$ action, according to the
adiabatic philosophy, preserves this fibration and acts as $\tau$ on
${\bf CP}^1$.  In particular, since $\tau$ acts freely on ${\bf CP}^1$,
$w$ acts freely on $\K3$.  Moreover, since the action of $w$ commutes with
half the supersymmetries, $w$ preserves one of the complex structures
of $\K3$ (though not the one in which the fibration $\K3\to {\bf CP}^1$
is holomorphic).  A ${\bf Z}_2$ symmetry of $\K3$ that preserves a complex
structure and acts freely is equivalent to the Enriques involution,
so we can identify $w$ with that involution, and we have determined
how $w$ acts on $\T^2\times \K3$.

Notice that the involution $w$ of $\T^2\times \K3$ is the same one used
on the Type II side in \fhsv\ to obtain a model with $N=2$ supersymmetry,
and therefore the $N=1 $ model we are constructing uses on the heterotic
string side the same Calabi-Yau manifold
 used in \fhsv\ on the Type II side.

To complete the description of the model on the heterotic string side,
we have to lift the action of $w$ to the $E_8\times E_8$ gauge bundle.
We cannot be as rigorous in describing this, but the fact that in the
construction
of \kachruvafa\ the second Chern class was equally divided between
the two $E_8$'s (otherwise complete $E_8\times E_8$ breaking was not possible)
creates an elegant possibility that would not exist otherwise:
$w$ can be the automorphism that exchanges the two $E_8$'s.
The same possibility exists in the example of \kachruvafa\ related
to the $X$ manifold since again the second Chern class is divided equally.

What happens to the hypermultiplets on the heterotic string side when
one divides by $w$? On general grounds, from each $N=2$ hypermultiplet
one $N=1$ chiral multiplet survives.  From the geometrical moduli
of $\K3$, the survivors are the moduli of the Enriques surface (twenty chiral
multiplets).
{}From the moduli of the    $E_8\times E_8$ bundle, assuming that $w$
exchanges the two $E_8$'s, what survive are the moduli associated with
one $E_8$ (224 chiral multiplets).  The other surviving chiral multiplets
are $S$, $T$, and $U$.

\subsec{Conifold Singularities And Transitions}

\nref\strom{A. Strominger, ``Massless Black Holes and Conifolds in
String Theory,'' hep-th/9504090.}
\nref\gms{B. Greene, D. Morrison, and A. Strominger, ``Black Hole
Condensation and the Unification of String Vacua,'' hep-th/9504145.}
One might wonder what becomes in such  $N=1$ orientifolds of
some of the interesting physics associated with conifold
singularities and their generalizations \refs{\strom-\gms}.

In $N=2$ models,
Type IIA vector multiplets do not couple to elementary charged
string states, but they do couple to charged black holes.
At the mirror of a conifold singularity, such a charged black
hole hypermultiplet goes to zero mass.   Let $M$ and $\tilde M$ be
the $N=1$ chiral superfields in such a hypermultiplet.
Let $A$ be a gauge field with respect to which $M$ and $\tilde M$
are charged, and let $a$ be the           $N=1$ chiral superfield
related to $A$ by $N=2$.  The couplings important at the conifold
singularity are a superpotential term $\Delta W = a M\tilde M$ and
the couplings of the gauge field to $M$ and $\tilde M$.
In conformal field theory, $M$ and $\tilde M$ have, in effect,
been integrated out, but two traces of this show up in couplings
of $a$ and $A$ computed in conformal field theory.
(1) There is a logarithmic  singularity in the Kahler metric on
the $a$ plane.  (2) The effective coupling of the $A$ field
(after performing a duality transformation to the ``right'' photon)
goes to zero at the conifold point, while there is a $2\pi$ monodromy
in the effective
$\theta$ angle as one winds around the conifold point.

Now if one goes to an orientifold $N=1$ model, what happens
in the weak coupling, conformal field theory limit is that
either $a$ or $A$ is projected out, but the couplings of the
field that does survive are unchanged.  If it is the chiral multiplet
$a$ that survives, then one has the logarithmic singularity
in the metric on the $a$ plane in the projected conformal field theory.
This is interpreted as resulting from integrating out   a chiral
superfield $M'$ with a superpotential $\Delta W=a(M')^2$.
Presumably $M'$ is a linear combination of $M$ and $\tilde M$.
\foot{It may be that both $M$ and $\tilde M$  survive.
In  an $N=1$ theory, given  a logarithmic singularity
$ds^2=|da|^2\ln |a|^2$ in the Kahler metric, there is no way
to infer how many chiral superfields with $a(M')^2$ couplings
were integrated out to produce the singularity.  This is because
the coefficient of the logarithm can be   absorbed in the normalization
of $a$.  For $N=2$, $a$ is in a supermultiplet with a gauge field
that has a natural normalization, and this gives a natural
normalization for $a$.}
In the other case, where $A$ survives, the vanishing effective
gauge coupling and monodromy in the $\theta$ angle that one sees
in the conformal field theory limit are signals that $A$ is coupled
to a charged multiplet that is becoming massless at the conifold
point.

Perhaps it should be stressed that the occurrence of these phenomena
-- though not the precise point in moduli space at which they occur --
is stable under quantum corrections to the conformal field theory
analysis.  That is because   these effects are associated with
certain topological invariants which cannot just disappear when
space-time quantum corrections
are turned on.  In the case that $A$ survives in the orientifold,
the topological invariant that makes the effect stable
 is the $2\pi$ shift of the $\theta$ angle in
a circuit around the locus  at which $M$ and $\tilde M$ are massless;
when $a$ survives, the relevant effect is a $2\pi$
shift in the $(M')^2$ (or $M\tilde M$) term in the
superpotential in a circuit around this locus; this shift  determines
a $2\pi$ shift in the phase of the mass of the fermion field in
$M'$. Note that topological
invariance of these $2\pi$ shifts in $\theta$ angles and phases
of fermion masses does not depend on supersymmtry, so massless charged
fermions associated with conifolds also survive in non-supersymmetric
orbifolds.

\bigskip
\noindent{\it Alternative Class Of $N=1$ Models}

The role of the conifold can be similarly discussed in
the  alternative class of $N=1$ models derived from Calabi-Yau manifolds,
in which one finds an involution $g$ that preserves the complex
structure and     holomorphic three-form, and divides
by $g(-1)^{F_L}$.  From an $N=2$ vector multiplet $V= (a,A)$,
the $N=1$ chiral multiplet $a$ survives in the projected theory if
$V$ is even under $g$, while the $N=1$ vector multiplet containing
$A$ survives if $V$ is odd under $g$.   The conformal field theory
couplings of the surviving modes are the couplings of the $N=2$ model,
and the rest of the discussion proceeds as above.
For this class of $N=1$ models, however, we cannot  reliably
exhibit heterotic string duals.

\subsec{Relation To Type I Models}

In this discussion, we have emphasized orientifolds of the Type IIA
model in which one divides by a symmetry that reverses orientation of
both the world-sheet and the target space-time.
The analog for Type IIB would be to divide by a symmetry that reverses
orientation of the world-sheet while preserving the orientation of the
target.  The two classes of models are closely related and are
very plausibly
 exchanged by mirror symmetry.

A special case of a Type IIB orientifold is the one obtained by dividing
by orientation reversal on the world-sheet together with trivial action on
space-time.  The $SO(32)$ Type I superstring should presumably be interpreted
as this orbifold of Type IIB.  As this example shows, in constructing
orientifolds, non-abelian gauge groups in space-time will naturally
arise from fixed points of the involution in space-time.

\nref\dabholkar{A. Dabholkar,
``Ten Dimensional Heterotic String as a Soliton,'' hep-th/9506160.}
\nref\chull{C. Hull, ``String-String Duality in Ten Dimensions,''
hep-th/9506194.}

It appears that many $N=1$ vacua of the heterotic string are equivalent
to Type II orientifolds.  For instance, it now seems \refs{\ewitten,\dabholkar
-\chull}\ that the $SO(32)$ heterotic
string in ten dimensions is equivalent to the Type I superstring in ten
dimensions.  So any heterotic string vacuum that can be understood as
a compactification of the $SO(32)$ model in ten dimensions will be equivalent
to an analogous Type I compactification. But our results
indicate that many other heterotic string vacua are equivalent to
orientifolds, including $E_8\times E_8$ models as well as the $SO(32)$
models that can be related directly to Type I.

We would like to acknowledge the hospitality of the International
Center for Theoretical Physics.
The research of C. Vafa is supported in part by NSF grant PHY-92-18167;
that of E. Witten, by NSF-PHY92-45317.

\listrefs
\end